\documentclass{article}

\usepackage[preprint]{neurips_2026}

\usepackage[utf8]{inputenc} %
\usepackage[T1]{fontenc}    %
\usepackage{amsmath,amssymb,amsthm}
\usepackage{mathtools}
\usepackage{url}            %
\usepackage{booktabs}       %
\usepackage{array}          %
\usepackage{amsfonts}       %
\usepackage{nicefrac}       %
\usepackage{microtype}      %
\usepackage{xcolor}         %
\usepackage{graphicx}
\usepackage{multirow}
\usepackage{cleveref}
\usepackage{wrapfig}
\usepackage{thmtools, thm-restate}
\usepackage[colorinlistoftodos]{todonotes} %
\usepackage{tikz}

\usetikzlibrary{calc,arrows.meta,positioning}
\usepackage[normalem]{ulem} %

\declaretheorem[style=plain, name=Theorem]{theorem}
\declaretheorem[style=plain, name=Proposition, sibling=theorem]{proposition}
\declaretheorem[style=plain, name=Lemma, sibling=theorem]{lemma}
\declaretheorem[style=plain, name=Corollary, sibling=theorem]{corollary}

\declaretheorem[style=definition, name=Definition, sibling=theorem]{definition}

\declaretheorem[style=remark, name=Remark, sibling=theorem]{remark}

\newenvironment{proofsketch}{\begin{proof}[Proof sketch]}{\end{proof}}

\newcommand{\R}{\mathbb{R}}
\newcommand{\E}{\mathbb{E}}

\title{Can Ideologues Agree on Quality? From Non-identifiable Latent Factors to Collective Outcomes}

\author{
  David Gamba \quad Seura Ha \quad Daniel M. Romero \quad Grant Schoenebeck \\
  University of Michigan \\
  \texttt{gamba@umich.edu, seura@umich.edu, drom@umich.edu, schoeneb@umich.edu}
}

\begin{document}

\maketitle

\begin{abstract}
To surface ``objective'' fact-check notes, X's Community Notes uses Biased Matrix Factorization (BMF) on crowd-sourced ratings to separate note quality from reviewer ideology. We show that for any correctly specified model that decomposes ratings into quality and ideological alignment, this separation requires fixing an ideological anchor: a reference point for what counts as neutral. We give a characterization for the standard implementation: with $L_2$ regularization, BMF anchors quality at a point proportional to the mean reviewer ideology, recovering objective quality if and only if the mean reviewer is neutral, assuming note quality is uncorrelated with note leaning. This parallels the Condorcet Jury Theorem: collective judgment recovers truth when the pool is unbiased on average. When that assumption fails, BMF's recovered ranking shifts toward maximizing aggregate reviewer satisfaction, which coincides with social welfare only when the pool is representative. This is not specific to BMF; the rating matrix admits a continuous family of gauge-equivalent decompositions, each implying a different quality ranking, so no mechanism based on ratings alone can recover quality without implicitly choosing such an anchor. What appears to be routine hyperparameter tuning is in fact a normative choice about whose preferences define quality.
\end{abstract}

\section{Introduction}
\label{sec:introduction}
 
When a post goes viral on X (formerly Twitter), a crowdsourced fact-checking note may appear beneath it. Whether that note is shown to millions of users is determined by an algorithm that at its core uses Biased Matrix Factorization (BMF), which aggregates ratings from volunteer reviewers to estimate the ``quality'' of each proposed note. The system, called Community Notes (formerly Birdwatch), has become one of the most prominent examples of algorithmic governance at scale \citep{wojcikBirdwatchCrowdWisdom2022, allenBirdsFeatherDont2022}. Similar systems are being adopted elsewhere: Meta has announced plans to replace professional fact-checkers with community-based evaluation \citep{vallanceMetaWantsXstyle2025, metaTestingBeginsCommunity2025}. As social media platforms have become major political arenas where algorithmic curation mediates discourse and often intensifies polarization \citep{pathakAnalyzingInterplayDiversity2024a, corsiEvaluatingTwittersAlgorithmic2023}, understanding these mechanisms becomes a pressing concern.
 
BMF's design aims to extract ``objective quality''\footnote{Community Notes documentation refers to ``idiosyncratic helpfulness.'' We interpret this as a quality measure used to decide whether to display a note.} from ideologically biased ratings \citep{wojcikBirdwatchCrowdWisdom2022}. The intuition is that if a note receives high ratings from both left-leaning and right-leaning reviewers, it is probably genuinely good, whereas a note praised only by one side may simply be ideologically aligned. BMF formalizes this by learning latent ideology vectors for both reviewers and notes, then defining quality as the residual, what persists after accounting for ideological alignment. \textbf{Whether this choice recovers objective quality is the central question of this paper.}
 
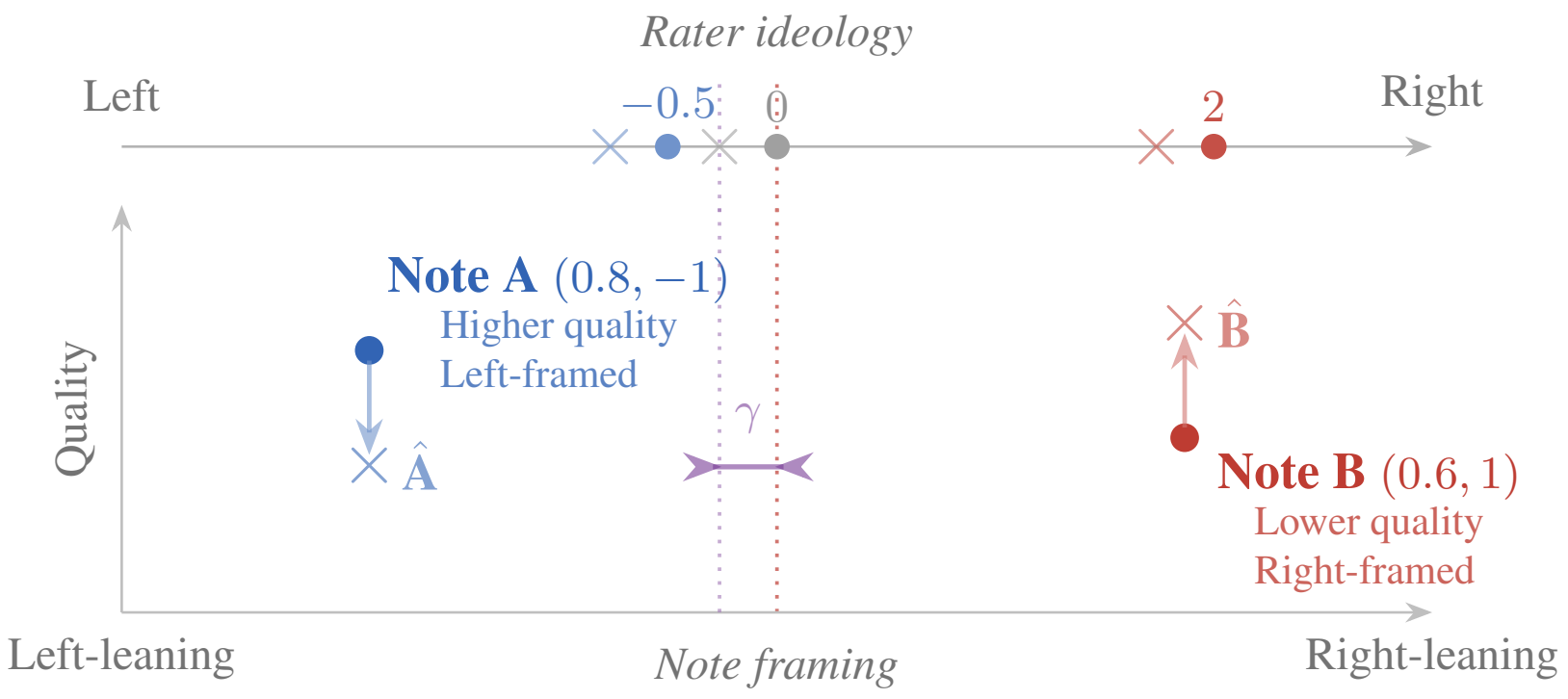
\begin{figure}[t]
    \centering
    % running_example_neurips.tex – version 1 (BMF γ, transparent overlay, dot/×, wider, ruler in middle)
\usetikzlibrary{calc,arrows.meta,positioning,shapes.geometric}
\begin{tikzpicture}[
    >=Stealth,
    every node/.style={font=\large},
    pref/.style={->, line width=0.7pt, shorten >=3pt, shorten <=2pt, opacity=0.45, dashed},
    shiftArrow/.style={->, line width=0.9pt, shorten >=2pt, shorten <=2pt, solid},
    axlabel/.style={font=\small, text=black!55},
    dot/.style={circle, fill, inner sep=0pt, minimum size=5pt},
    cross/.style={font=\Large\bfseries}
]

\definecolor{leftblue}{RGB}{50,100,180}
\definecolor{rightred}{RGB}{190,60,50}
\definecolor{centergray}{RGB}{120,120,120}
\definecolor{noteA}{RGB}{50,100,180}
\definecolor{noteB}{RGB}{190,60,50}
\definecolor{gammaCol}{RGB}{120,60,150}

% ===== Ground truth (full opacity) =====
\begin{scope}
  % Neutral reference line at x=0
  \draw[rightred!70, line width=0.7pt, dotted] (0, 2.75) -- (0, 3.65);
  \draw[rightred!70, line width=0.7pt, dotted] (0, 0.0)  -- (0, 2.8);
  % Rater ideology axis (widened)
  \draw[->, black!30, line width=0.5pt] (-4.5, 3.2) -- (4.5, 3.2);
  \node[axlabel] at (-4.5, 3.55) {Left};
  \node[axlabel] at (4.5, 3.55) {Right};
  % Increased headroom for the axis label
  \node[axlabel, above=0.5cm, font=\small\itshape] at (0, 3.2) {Rater ideology};

  % Original users: dot with number above
  \node[dot, draw=none, fill=leftblue!70] (a1) at (-0.75, 3.2) {};
  \node[above=0.05cm, font=\small\bfseries, text=leftblue!80] at (a1) {$-0.5$};
  \node[dot, draw=none, fill=centergray!70] (a2) at (0, 3.2) {};
  \node[above=0.05cm, font=\small\bfseries, text=centergray!80] at (a2) {$0$};
  \node[dot, draw=none, fill=rightred!90] (a5) at (3.0, 3.2) {};
  \node[above=0.05cm, font=\small\bfseries, text=rightred!90] at (a5) {$2$};

  % Note plane axes (widened)
  \draw[->, black!25, line width=0.5pt] (-4.5, 0.0) -- (4.5, 0.0);
  \draw[->, black!25, line width=0.5pt] (-4.5, 0.0) -- (-4.5, 2.8);
  \node[axlabel] at (4.5, -0.32) {Right-leaning};
  \node[axlabel] at (-4.5, -0.32) {Left-leaning};
  \node[axlabel, rotate=90, anchor=south] at (-4.55, 1.4) {Quality};
  \node[axlabel, below=0.12cm, font=\small\itshape] at (0, 0.0) {Note framing};

  % Original notes (circles) with coordinates
  \fill[noteA] (-2.8, 1.8) circle (2.8pt);
  \node[font=\normalsize\bfseries, text=noteA,
        above right=0.2cm and 0.0cm of {(-2.8,1.8)}] (nA) {Note A \small{$(0.8,-1)$}};
  \node[font=\footnotesize, text=noteA!80,
        below=-0.2cm of nA, anchor=north, align=left]
      {Higher quality\\Left-framed};
  \fill[noteB] (2.8, 1.2) circle (2.8pt);
  \node[font=\normalsize\bfseries, text=noteB,
        below right=-0.02cm and 0.1cm of {(2.8,1.2)}] (nB) {Note B \small{$(0.6,1)$}};
  \node[font=\footnotesize, text=noteB!80,
        below=-0.2cm of nB, anchor=north, align=left]
      {Lower quality\\Right-framed};

  % Preference arrows (commented out)
  % \draw[pref, leftblue] ...
\end{scope}

% ===== Transformed (γ=0.263 → shift = -0.3945 cm) =====
\begin{scope}[opacity=0.6]
  \def\gammaX{-0.3945}

  % Shifted neutral reference (dotted line)
  \draw[gammaCol!70, line width=0.7pt, dotted] (\gammaX, 2.75) -- (\gammaX, 3.65);
  \draw[gammaCol!70, line width=0.7pt, dotted] (\gammaX, 0.0)  -- (\gammaX, 2.8);

  % Gamma measurement ruler: placed in the middle, between the two axes
  \draw[>-<, >=Stealth, gammaCol, line width=1pt]
    (\gammaX-0.25, 1.0) -- (0+0.25, 1.0)
    node[midway, above=0.1cm, font=\small\itshape, text=gammaCol] {$\gamma$};

  % Shifted agents: × markers
  \node[cross, text=leftblue!70] (a1t) at ({-0.75+\gammaX}, 3.2) {$\times$};
  \node[cross, text=centergray!70] (a2t) at ({0+\gammaX}, 3.2) {$\times$};
  \node[cross, text=rightred!90] (a5t) at ({3.0+\gammaX}, 3.2) {$\times$};

  % Shifted quality: × marks
  \node[cross, text=noteA] at (-2.8, 1.011) {$\times$};
  \node[font=\normalsize\bfseries, text=noteA,
        right=0.1cm of {(-2.8,1.011)}] {$\hat{\text{A}}$};
  \node[cross, text=noteB] at (2.8, 1.989) {$\times$};
  \node[font=\normalsize\bfseries, text=noteB,
        right=0.1cm of {(2.8,1.989)}] {$\hat{\text{B}}$};

  \draw[shiftArrow, noteA!70] (-2.8, 1.8) -- (-2.8, 1.011);
  \draw[shiftArrow, noteB!70] (2.8, 1.2) -- (2.8, 1.989);

  % Shifted preference arrows (commented out)
  % \draw[pref, leftblue] ...
\end{scope}
\end{tikzpicture}
    \caption{Running example. Three reviewers ($u \in \{-0.5,0,2\}$, dots) evaluate Note~A (left-leaning, higher-quality) and Note~B (right-leaning, lower-quality). Crosses show the gauge-shifted representation; %
    reviewer ideologies shift left by $\gamma$ while note qualities adjust proportionally, leaving all ratings unchanged but reversing the ranking ($\hat{q}_B > \hat{q}_A$). BMF implicitly selects this shifted frame.}
    \label{fig:running_example}
\end{figure}
 
To answer this question, we assume ratings are generated by a ground truth model in which BMF is correctly specified. That is, each reviewer $i$ has a latent ideology $u_i \in \R$ and each note $j$ has a true quality $q_j \in \R$ and an ideological leaning $v_j \in \R$, with ratings given by:
\begin{equation}
    r_{ij} = q_j + u_i \, v_j + \varepsilon_{ij},
    \label{eq:ground_truth}
\end{equation}
 
where $\varepsilon_{ij}$ are i.i.d.\ noise terms. This matches the functional form that BMF uses to model rating generation (\Cref{eq:bmf_model}), with `true' parameters in place of estimated ones. For the algorithm analysis, we take the noiseless limit: if BMF fails to recover quality here, it cannot do so in more general conditions.

Our analysis characterizes precisely what BMF recovers, starting from a simple observation: the decomposition of a rating into ``quality'' and ``ideological alignment'' is not unique. One can shift all reviewer ideologies by a constant and compensate by adjusting quality scores, without changing any observed or predicted rating (as illustrated in \Cref{fig:running_example}). This \emph{gauge symmetry}, or \emph{non-identifiability}, means the data alone cannot determine which reviewers are neutral and which are biased; different anchor choices lead to different quality rankings. We show that for extracting quality from ratings, BMF implicitly designates some ideology as the reference point---an \emph{anchor} defining which perspective is ``neutral''. More specifically, we find that regularization plays a major role in this:

\begin{enumerate}
    \item \textbf{$L_2$ regularization anchors at the mean ideology.} When BMF resolves the gauge symmetry ambiguity via $L_2$ regularization, the recovered qualities satisfy $\hat{q}_j = q_j + \gamma^* v_j$ up to a constant, where $\gamma^* = s\,\mu_u$, $\mu_u$ is the mean reviewer ideology, and $s \in (0,1)$ (\Cref{thm:anchor_l2}). Quality is recovered when $\mu_u = 0$; otherwise rankings are shifted toward the average reviewer's perspective (\Cref{cor:quality_recovery_full}).
    \item \textbf{No mechanism on ratings alone can do better.} For any two notes with different ideological leanings, there exists a gauge-equivalent parameterization that preserves all ratings but reverses their quality ordering (\Cref{thm:quality-nonidentifiable}). This is a property of the rating data itself: \emph{no} mechanism operating on ratings alone can recover quality (\Cref{cor:no_mechanism}).
\end{enumerate}

Just as the Condorcet Jury Theorem guarantees majority vote recovers truth when the voter pool is unbiased on average \citep{condorcetEssaiLapplicationLanalyse1785}, BMF recovers quality exactly when the reviewer pool is ideologically balanced ($\mu_u = 0$), assuming note quality is uncorrelated with note leaning. This limitation is not an artifact of BMF's particular algorithm but a property of the rating data itself, so no mechanism can escape it. What appears to be a routine choice of regularization weights is in fact a normative decision about whose ideological perspective defines quality; we discuss implications for platform design in \Cref{sec:discussion}.

\paragraph{Related work.}
BMF is a standard recommender-systems technique \citep{korenMatrixFactorizationTechniques2009, bellLessonsNetflixPrize2007, heNeuralCollaborativeFiltering2017} adapted for content evaluation in Community Notes \citep{wojcikBirdwatchCrowdWisdom2022}. Our model \eqref{eq:ground_truth} is closely related to ideal-point models in political science \citep{pooleSpatialModelLegislative1985, clintonStatisticalAnalysisRoll2004} and item-response theory in psychometrics, both of which have long required explicit anchoring for interpretation. Non-identifiability in factor models has been studied since \citet{andersonStatisticalInferenceFactor1956, baiStatisticalAnalysisFactor2012}; conditions for identifiable representation learning remain an active area \citep{ahujaPropertiesMechanismsEquivariance2021, locatelloChallengingCommonAssumptions2019, parkLinearRepresentationHypothesis2024, zouRepresentationEngineeringTopDown2025}. Our result parallels \citet{locatelloChallengingCommonAssumptions2019} and \citet{damourMultiCauseCausalInference2019}: just as disentanglement is impossible without inductive biases and latent confounders are not identified by treatment structure alone, quality recovery is impossible without an ideological anchor.

Content moderation can also be viewed as a social choice problem \citep{robertsScreenContentModeration2019, wangEfficiencyCommunityBasedContent2024}, connecting to classical and computational social choice \citep{arrowSocialChoiceIndividual2012, brandtIntroductionComputationalSocial2016, boutilierOptimalSocialChoice2015} and to recent work framing ML systems as preference aggregators \citep{qiuRepresentativeSocialChoice2025, siththaranjanDistributionalPreferenceLearning2024, fishGenerativeSocialChoice2023, christianoDeepReinforcementLearning2023}, which face well-known strategic vulnerabilities \citep{gibbardManipulationVotingSchemes1973}. The tension between aggregation and fairness in recommender systems \citep{ekstrandAllCoolKids2018, geExplainableFairnessRecommendation2022, dworkFairnessAwareness2012}, together with calls for algorithmic accountability \citep{krollAccountableAlgorithms2017, mitchellModelCardsModel2019}, motivates making such implicit normative choices explicit.

\section{Biased Matrix Factorization and the Correctly Specified Setting}
\label{sec:setup}

\subsection{The BMF Mechanism}

We consider $n$ reviewers indexed by $i \in [n]$ and $m$ notes indexed by $j \in [m]$. The rating of reviewer $i$ on note $j$ is $r_{ij} \in \R$ (with $r_{ij} = \bot$ if unobserved). 

\begin{definition}[Biased Matrix Factorization]
\label{def:bmf}
BMF models the rating of reviewer $i$ on note $j$ as:
\begin{equation}
    \hat{r}_{ij} = \hat{c} + \hat{b}_i + \hat{q}_j + \hat{u}_i \cdot \hat{v}_j,
    \label{eq:bmf_model}
\end{equation}
where $\hat{c} \in \R$ is a global intercept, $\hat{b}_i \in \R$ is a reviewer-specific intercept, $\hat{q}_j \in \R$ is interpreted as the quality of note $j$, and $\hat{u}_i, \hat{v}_j \in \R^d$ are latent ideology factors. Parameters are estimated by minimizing:
\begin{equation}
    \mathcal{L}(\hat\theta) = R(\hat\theta) + \Lambda(\hat\theta) \;= \sum_{\substack{i,j:\, r_{ij} \neq \bot}} (r_{ij} - \hat{r}_{ij})^2 + \Lambda(\hat\theta),
    \label{eq:bmf_objective}
\end{equation}
where $\hat\theta=\{\hat c, \hat b_i, \hat q_j, \hat u_i, \hat v_j\}$, $R(\hat\theta)$ is the empirical risk and $\Lambda(\hat\theta)$ the regularization term.
\end{definition}

We focus on the model with scalar ideology factors ($d=1$), matching Community Notes' implementation. The $d>1$ case introduces one latent dimension per ideological axis; the non-identifiability results of \Cref{sec:gauge,sec:tightness} extend, since each axis admits a shift ambiguity of the kind formalized in \Cref{sec:gauge}, while the anchor characterization of \Cref{sec:regularization} is specific to $d = 1$ (\Cref{sec:discussion}). The key interpretive claim is that $\hat{q}_j$ captures intrinsic quality independent of ideology; notes are selected for display when $\hat{q}_j$ exceeds a threshold.

We note that Community Notes uses categorical ratings (helpful/not helpful/somewhat helpful) rather than continuous scores, and employs logistic loss rather than squared error. Our analysis works in the underlying \emph{score space} (given by continuous $\hat r_{i,j}$), where BMF fits parameters in both cases; the logistic link $\sigma(\cdot)$ enters only when scores are mapped to rating probabilities for the categorical loss, leaving the structure we study unchanged. Community Notes also includes additional filtering steps (minimum rating counts, tag-based weighting) that we do not model. Our analysis focuses on the core factorization mechanism, and our empirical results (\Cref{sec:experiments}) verify that the main findings are robust when extended to a binary setting (helpful/not helpful).
 
\subsection{The Correctly Specified Setting}

To evaluate whether BMF recovers objective quality, we need a setting with well-defined ground truth. We assume ratings are generated by a model that matches BMF's functional form (\Cref{eq:bmf_model}), with the same global intercept $c$, per-reviewer intercepts $b_i$, quality terms $q_j$ which represent `objective' quality, and ideological alignment $u_i v_j$. If reviewer and note share the same ideological sign, the reviewer rates it above quality alone.

\begin{definition}[Ground Truth Model]
\label{def:ground_truth}
Each reviewer $i$ has a true ideology $u_i \in \R$ and intercept $b_i \in \R$. Each note $j$ has a true quality $q_j \in \R$ and ideological leaning $v_j \in \R$. With a global intercept $c \in \R$, ratings are generated as:
\begin{equation}
    r_{ij} = c + b_i + q_j + u_i \, v_j + \varepsilon_{ij},
    \label{eq:dgp}
\end{equation}
where $\varepsilon_{ij}$ are i.i.d.\ noise terms with mean zero and variance $\Delta$.
\end{definition}

For the analysis of what BMF recovers in \Cref{sec:regularization} we additionally set $c = b_i = 0$, which we call the \emph{simplified ground truth model}, isolating the two elements central to our question: note quality $q_j$ and ideological alignment $u_i v_j$. This simplifies the proof for \Cref{thm:anchor_l2} while keeping the interaction between quality and ideological alignment; the result extends under weaker conditions on $c$ and $b_i$ (\Cref{app:proofs:regularization}). The gauge and tightness results in \Cref{sec:gauge,sec:tightness} hold for the full ground truth model without this simplification.  

\begin{wrapfigure}[20]{l}{0.5\linewidth}
\includegraphics[width=0.97\linewidth]{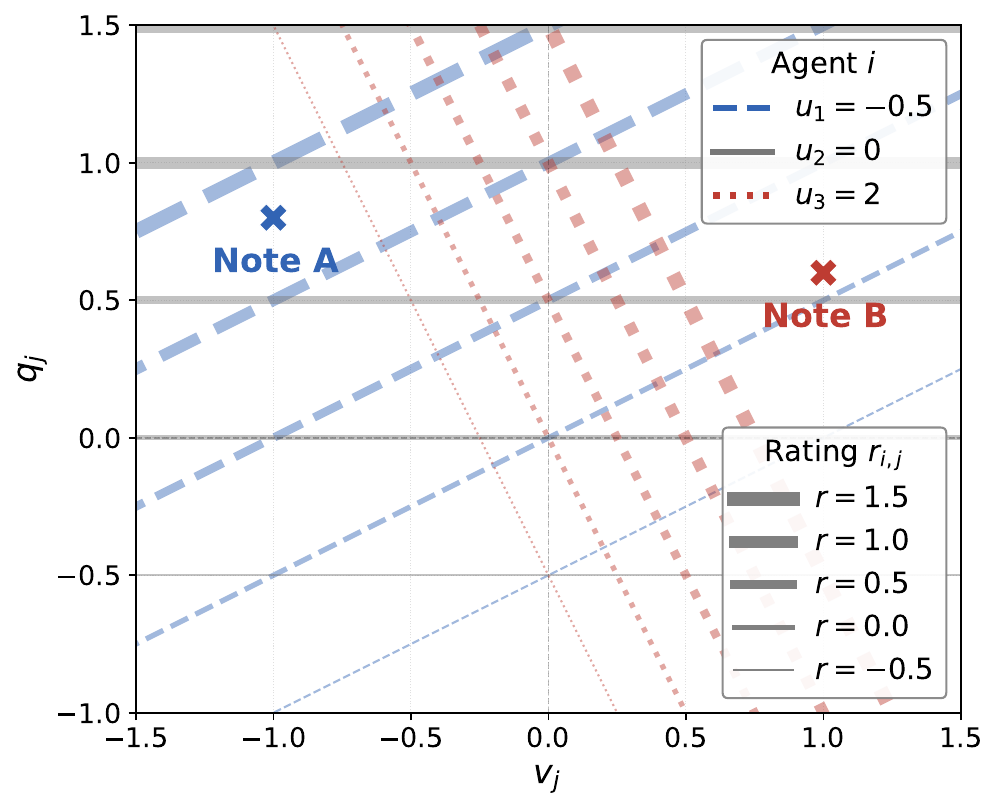}
\caption{Iso-rating contours for the running example (\Cref{fig:running_example}). Contour lines show constant rating for reviewers. %
The neutral reviewer's contours are horizontal (rating equals quality); ideological reviewers have tilted contours, systematically over- or under-rating ideologically aligned notes.}
\label{fig:iso_utility_lines}
\end{wrapfigure}

We analyze the noiseless setting ($\Delta = 0$) under correct specification, the most favorable setting for the algorithm: if BMF fails to recover quality on data its own model generated, noise does not help it (\Cref{cor:noisy_persistence}), and misspecification did not help it in the three departures we tested (\Cref{app:experiments:misspecification}). On real ratings the true ideologies and qualities are unobserved, so there is no standard against which to check the anchor; the synthetic setting supplies one.
 
A reviewer with $u_i = 0$ is \emph{ideologically neutral}; their ratings rank notes by quality. This leads to a key observation:

\begin{proposition}[Neutral-reviewer ratings rank by quality]
\label{prop:neutral_agent}
In the ground truth model, for any reviewer $i$ with $u_i = 0$ and any two notes $j_1, j_2$: $q_{j_1} > q_{j_2}$ if and only if $\E[r_{ij_1}] > \E[r_{ij_2}]$.
\end{proposition}

\begin{proof}
$\E[r_{ij} \mid u_i = 0] = c + b_i + q_j$. Since $c + b_i$ is constant across notes, $\E[r_{ij_1}] - \E[r_{ij_2}] = q_{j_1} - q_{j_2}$.
\end{proof}
 
Thus, recovering the ranking of items according to objective quality is equivalent to recovering the perspective of the neutral reviewer. The question is whether BMF can identify which perspective is neutral from ratings alone.
 
\paragraph{Running example (\Cref{fig:running_example}).}  Throughout, the running example uses the simplified model setting ($c = b_i = 0$). Two candidate fact-checking notes have been proposed: Note~A is higher quality with left-leaning framing ($q_A = 0.8$, $v_A = -1$); Note~B is of lower quality but right-leaning ($q_B = 0.6$, $v_B = 1$). Three reviewers span the ideological spectrum: $u_1 = -0.5$ (left), $u_2 = 0$ (neutral), $u_3 = 2$ (far-right). The pool skews right (average ideology $0.5$). Ratings are tabulated in \cref{app:running_example}.

\subsection{Gauge Symmetries of BMF}
\label{sec:gauge}
 
The running example illustrates a key property of the rating model: shifting all reviewer ideologies by a constant while adjusting qualities to compensate leaves every rating unchanged. We characterize the full family of such transformations and identify which one affects quality rankings. %

\begin{lemma}[Gauge Invariance of Ratings]
\label{lem:gauge_invariance}
In the ground truth model, for any $\gamma \in \R$, define $u_i' = u_i - \gamma$, $q_j' = q_j + \gamma v_j$, $v_j' = v_j$. Then for all $i, j$: $r_{ij}'= c + b_i + q_j' + u_i'v_j'= r_{ij}$.
\end{lemma}
 
\begin{proof}
 $c + b_i + q_j' + u_i' v_j' = c + b_i + (q_j + \gamma v_j) + (u_i - \gamma) v_j = c + b_i + q_j + u_i v_j$.
\end{proof}
 
The two parameter sets generate identical ratings for all reviewer-note pairs. The term ``gauge invariance'' is borrowed from physics, where it describes symmetries under which observable quantities are unchanged despite changes in representation. Here, ratings are observable; latent factors are not.

This symmetry extends to the full BMF model \eqref{eq:bmf_model}, which admits a richer family of transformations. We characterize all of them, then identify which one matters for quality recovery.
 
\begin{theorem}[restate=thmGaugeExhaustive, 
                name=Characterization of BMF's Gauge Symmetries]
\label{thm:gauge_exhaustive}
Every gauge transformation of BMF (i.e., every parameter mapping that preserves all predicted ratings $\hat{r}_{ij}$) can be parameterized by $(\alpha, \beta, \gamma, \eta, \tau)$ with $\alpha, \beta, \gamma, \eta \in \R$ and $\tau \in \R \setminus \{0\}$:
\begin{equation}
    \begin{gathered}
        \hat{u}_i' = \tau(\hat{u}_i - \gamma), \qquad
        \hat{q}_j' = \hat{q}_j + \gamma \hat{v}_j + \beta, \\[4pt]
        \hat{v}_j' = \hat{v}_j / \tau - \eta / \tau, \qquad
        \hat{b}_i' = \hat{b}_i + \hat{u}_i \eta + \alpha, \qquad
        \hat{c}' = \hat{c} - \alpha - \beta - \gamma\eta.
    \end{gathered}
    \label{eq:full_gauge}
\end{equation}
These five parameters exhaust all degrees of non-identifiability.
\end{theorem}

\begin{proofsketch}
A derivative argument on the bilinear constraint $\hat{r}_{ij}' = \hat{r}_{ij}$ shows that rating-preserving maps are affine in each factor. Requiring the same form to hold for every reviewer-note pair forces coefficients to be index-independent, leaving five universal scalars.
\end{proofsketch}

Each of the five parameters in \eqref{eq:full_gauge} corresponds to a distinct rating-preserving symmetry. $\gamma$ is the agent-centric shift of \Cref{lem:gauge_invariance}, moving the ideological zero among reviewers; $\eta$ is its item-side analogue, a shift in note ideology absorbed into the per-reviewer intercept; $\tau$ rescales reviewer and note ideologies reciprocally, leaving the bilinear product $\hat{u}_i \hat{v}_j$ unchanged; and $\alpha, \beta$ are additive offsets that redistribute constants among $\hat{c}, \hat{b}_i, \hat{q}_j$. Among these five symmetries, only one affects the quality ranking. %
 
\begin{proposition}[restate=propGammaOrdering, name=Only $\gamma$ Affects Quality Ordering]
\label{prop:gamma_ordering}
The relative ordering of $\hat{q}_j$ is invariant under the item-centric shift ($\eta$), scaling ($\tau$), and global offsets ($\alpha, \beta$). Only the agent-centric shift $\gamma$ can change the ordering.
\end{proposition}
 
\begin{proofsketch}
From \eqref{eq:full_gauge}, $\hat{q}_j' = \hat{q}_j + \gamma \hat{v}_j + \beta$. For two notes: $\hat{q}_{j_1}' - \hat{q}_{j_2}' = (\hat{q}_{j_1} - \hat{q}_{j_2}) + \gamma(\hat{v}_{j_1} - \hat{v}_{j_2})$. Only $\gamma$ appears in this difference; $\beta$ cancels. The item transformation changes $\hat{v}_j$ but not $\hat{q}_j$; scaling by $\tau$ preserves the inner product $\hat{u}_i \hat{v}_j$ and does not enter $\hat{q}_j$ directly. Full proof in \Cref{app:proofs:gauge}.
\end{proofsketch}

In the running example, choosing $\gamma = 1$ shifts the neutral reference from $u=0$ to $u=1$: the quality estimates become $\hat{q}_A = -0.20$ and $\hat{q}_B = 1.60$, re-ranking Note~B above Note~A despite its lower true quality, while leaving every entry of the rating matrix unchanged (\Cref{tab:ratings:shift} in \Cref{app:running_example}).
 
This result simplifies the analysis considerably: for quality rankings, the only consequential degree of freedom is $\gamma$, i.e., which ideology is treated as neutral. By \Cref{prop:neutral_agent}, this is the ideology whose perspective ranks notes by quality. The remaining gauge parameters either cancel in pairwise quality differences or are absorbed by the constant offset $\beta$. The question then becomes: how does BMF, in practice, choose $\gamma$?

\section{How $L_2$ Regularization Resolves the Ambiguity}
\label{sec:regularization}

We have established that the risk component $R$ of BMF's loss is gauge-invariant, so its minimizers form a manifold of equivalent solutions. In practice, however, BMF returns a single answer because the regularizer resolves this ambiguity. \Cref{prop:gamma_ordering} reduces BMF's ranking ambiguity to a single scalar $\gamma$. We now derive where $L_2$ regularization places it.
 
The intuition is as follows. $L_2$ regularization balances two forces: penalizing $\sum_i \hat{u}_i^2$ favors centering reviewer ideologies at zero, which pulls the anchor $\gamma$ toward the population mean $\mu_u$; penalizing $\sum_j \hat{q}_j^2$ favors small quality estimates, which pulls $\gamma$ toward zero. The ratio $\kappa = \lambda_q/\lambda_u$ controls this trade-off. In the remainder of this section, we formalize this intuition through a set of balance equations and solve for the resulting anchor.

We will show (\Cref{thm:anchor_l2}) that the selected anchor is $\gamma^* = s\,\mu_u$ for some $s \in (0,1)$ depending on regularization weights. This implies our main result: BMF recovers quality only if $\mu_u = 0$.

\paragraph{Notation.} We write $\mu_u = \tfrac{1}{n}\sum_i u_i$ and $\mu_v = \tfrac{1}{m}\sum_j v_j$ for the empirical means of reviewer and note ideologies, $\sigma_u^2 = \tfrac{1}{n}\sum_i (u_i - \mu_u)^2$ and $\sigma_v^2 = \tfrac{1}{m}\sum_j v_j^2$ for the corresponding empirical variances ($\sigma_v^2$ is uncentered; it is the variance when $\mu_v = 0$, as \Cref{thm:anchor_l2} assumes), and $\rho = m/n$ for the note-to-reviewer ratio. Analogous shorthands $\mu_q$, $\mu_b$ are used as needed.
 
\subsection{The Min-Min Solution}

We begin by formalizing how regularization interacts with the gauge orbit. Recall that $\Delta$ denotes the noise variance from \Cref{def:ground_truth}. In the noiseless case ($\Delta = 0$), the true parameters achieve zero reconstruction loss. We write $\mathcal{G}(\theta_{\mathrm{true}})$ for the \emph{gauge orbit} of the true parameters --- the set of all parameter vectors producing the same predictions as $\theta_{\mathrm{true}}$:
\[
    \mathcal{G}(\theta_{\mathrm{true}}) = \{ T(\theta_{\mathrm{true}}) : T \text{ is a gauge transformation \eqref{eq:full_gauge}} \}.
\]
This is the set of all zero-risk parameterizations. Regularization selects from this set by preferring ``simpler'' solutions.
 
\begin{definition}[Min-Min Solution]
\label{def:minmin}
The \emph{min-min solution} $\theta^*_\Lambda$ is the element of $\mathcal{G}(\theta_{\mathrm{true}})$ minimizing $\Lambda(\theta)$ (unique under $L_2$, \Cref{lem:unique_reduced}). The reflection $(\hat{u}, \hat{v}) \mapsto (-\hat{u}, -\hat{v})$, the gauge element $\tau = -1$, leaves $R$, $\Lambda_{L_2}$, and $\hat{q}$ unchanged; we identify parameter vectors that differ by it, which amounts to taking $\tau > 0$ on the orbit. A starred quantity for a parameter ($\gamma^*, \tau^*, \alpha^*, \dots$) denotes its value at $\theta^*_\Lambda$.
\end{definition}

\begin{proposition}[Regularized BMF converges to the min-min solution]
\label{prop:minmin_convergence}
In the noiseless setting, write the regularizer as $\lambda\Lambda(\theta)$, with $\lambda > 0$ a common factor and the ratios of $\lambda_u, \lambda_q, \lambda_v, \lambda_b, \lambda_c$ fixed. As $\lambda \to 0$, the minimizer $\theta_\lambda$ of $R(\theta) + \lambda \Lambda(\theta)$ converges to $\theta^*_\Lambda$.
\end{proposition}
 
\begin{proof}
By optimality of $\theta_\lambda$ against $\theta^*_\Lambda$ (which has zero risk): $R(\theta_\lambda) + \lambda \Lambda(\theta_\lambda) \leq \lambda \Lambda(\theta^*_\Lambda)$. Hence $R(\theta_\lambda) \to 0$ as $\lambda \to 0$, forcing $\theta_\lambda$ onto the zero-risk manifold $\mathcal{G}(\theta_{\mathrm{true}})$. Among zero-risk solutions, $\theta^*_\Lambda$ uniquely minimizes $\Lambda$ (\Cref{lem:unique_reduced}). By standard results in perturbation theory, the minimizer of a penalized objective converges to the minimizer of the penalty on the limiting constraint set as the penalty weight goes to zero. Hence $\theta_\lambda \to \theta^*_\Lambda$.
\end{proof}
 
\subsection{Balance Equations}  

The min-min solution must satisfy first-order optimality conditions along each gauge direction. We call these \emph{balance equations}. Since the risk is constant on the gauge orbit, the gradient of the regularizer must vanish along each symmetry direction. All five balance equations hold; the intercept and item-centric ones can be shown to force $\hat{b}_i, \hat{c} \to 0$ under the simplified ground truth model ($c = b_i = 0$). The two equations that determine $\gamma^*$ and hence quality rankings are the agent-centric balance (by \Cref{prop:gamma_ordering}) and the scale balance, which couples with it.

We consider $L_2$ regularization of the form
\begin{equation}
    \Lambda_{L_2}(\theta) = \lambda_u \sum_i \hat{u}_i^2 + \lambda_q \sum_j \hat{q}_j^2 + \lambda_v \sum_j \hat{v}_j^2 + \lambda_b \sum_i \hat{b}_i^2 + \lambda_c \hat{c}^2.
\end{equation}
 
\begin{lemma}[restate=lemBalanceLtwo, name=Balance Equations for $L_2$]
\label{lem:balance_l2}
Any minimizer of the $L_2$-regularized BMF objective satisfies:
\begin{equation}\label{eq:balance}
\begin{aligned}
\lambda_u \textstyle\sum_i \hat{u}_i = \lambda_q \textstyle\sum_j \hat{q}_j\hat{v}_j \quad &(\text{Agent-centric}) &\quad
\lambda_u \textstyle\sum_i \hat{u}_i^2 = \lambda_v \textstyle\sum_j \hat{v}_j^2 \quad &(\text{Scale}) \\
\lambda_b \textstyle\sum_i \hat{b}_i = \lambda_q \textstyle\sum_j \hat{q}_j = \lambda_c \hat{c} \quad &(\text{Intercept}) &\quad
\lambda_b \textstyle\sum_i \hat{b}_i\hat{u}_i = \lambda_v \textstyle\sum_j \hat{v}_j \quad &(\text{Item-centric})
\end{aligned}
\end{equation}
\end{lemma}
\begin{proofsketch}
For each gauge symmetry, the risk is invariant along the corresponding orbit, so the regularization term $\Lambda$ must be stationary at the minimum in each gauge direction. Each equation follows by differentiating $\Lambda$ with respect to the corresponding gauge parameter and setting the derivative to zero. Full proof in \Cref{app:proofs:regularization}.
\end{proofsketch}

The agent-centric balance has a natural interpretation: if $\sum_j \hat{q}_j \hat{v}_j = 0$ (quality and ideology uncorrelated), it  requires $\sum_i \hat{u}_i = 0$, centering reviewer ideologies at zero. The right-hand side pulls the anchor away from this centering by an amount proportional to the quality-ideology covariance. The scale equation balances the norms of reviewer and note ideology vectors; together with the agent-centric it forms a coupled system whose unique solution is $(\gamma^*, \tau^*)$.
 
\subsection{Main Result: BMF Anchors Proportional to the Mean}

We now characterize the anchor selected by $L_2$ regularization. The main result is that the anchor lands at a fraction of the mean reviewer ideology: $\gamma^* = s\mu_u$ for some $s \in (0,1)$.%

\begin{theorem}[restate=thmAnchorLtwo, name=BMF Anchors proportional to $\mu_u$]
\label{thm:anchor_l2}
Consider ratings generated by the simplified ground truth model ($c = b_i = 0$\footnote{The hypothesis $c = b_i = 0$ can be relaxed to $c + \mu_b = 0$ and $\mathrm{Cov}(b_i, u_i) = 0$ (\Cref{app:proofs:regularization}); we keep the simplified form in the main statement to streamline Step~1 of the proof sketch.}, \Cref{def:ground_truth}) with mean reviewer ideology $\mu_u$ and variance $\sigma_u^2 > 0$, zero-mean notes ($\mu_{q} = \mu_{v} = 0$), and $\mathrm{Cov}(q_j, v_j) = 0$. Under $L_2$ regularization, as $\lambda \to 0$, BMF recovers $\hat{q}_j = q_j + \gamma^* v_j + \beta$ where $\beta$ is a ranking-irrelevant constant and
\begin{equation}
    \gamma^* = \frac{\mu_u}{1 + \kappa \rho \sigma_v^2 / \tau^{*2}}, \qquad
    \tau^{*4} = \frac{\lambda_v m \sigma_v^2}{\lambda_u n [\sigma_u^2 + (\mu_u - \gamma^*)^2]},
    \label{eq:gamma_star}
\end{equation}
with $\kappa = \lambda_q/\lambda_u$ and $\rho = m/n$. In particular, $\gamma^* = s \mu_u$ for some $s \in (0,1)$.
\end{theorem}

\begin{proofsketch}
The proof has three steps:

\textbf{Step 1: Reduce to two unknowns.} By \Cref{prop:minmin_convergence}, the solution converges to the min-min solution, which minimizes $\Lambda_{L_2}$ over the gauge orbit $\mathcal{G}(\theta_{\mathrm{true}})$. On this orbit, the agent-centric shift transforms parameters as $\hat{u}_i = \tau(u_i - \gamma)$ and $\hat{q}_j = q_j + \gamma v_j$. The item-centric shift $\eta$ and global offsets $\alpha, \beta$ can be shown to vanish at the minimum: for fixed $(\gamma, \tau)$, the regularizer is strictly convex and quadratic in $(\eta, \alpha, \beta)$, and the moment conditions $\mu_q = \mu_v = 0$ eliminate the linear terms, forcing $\eta^* = \alpha^* = \beta^* = 0$ (\Cref{app:proofs:regularization}, Step~2).

This reduces the problem to two equations in $(\gamma, \tau)$. 

\textbf{Step 2: Apply the balance equations.} The two remaining free parameters $(\gamma, \tau)$ must satisfy the agent-centric and scale balance equations (\Cref{lem:balance_l2}). Substituting $\hat{u}_i = \tau(u_i - \gamma)$, $\hat{q}_j = q_j + \gamma v_j$ into the agent-centric balance, using $\mathrm{Cov}(q,v) = 0$, and solving for $\gamma$ yields $\gamma^*$ in \eqref{eq:gamma_star}. Similarly, the scale balance yields an expression for $\tau^*$.

\textbf{Step 3: $s \in (0,1)$.} The coupled system reduces to a single monotone equation in $\tau^2$ with exactly one root, so the fixed point $(\gamma^*, \tau^*)$ is unique (\Cref{lem:unique_reduced}). Since $\kappa, \rho, \sigma_v^2, \tau^{*2} > 0$ ensure the denominator in \eqref{eq:gamma_star} exceeds 1, $\gamma^* = s\mu_u$ with $s \in (0,1)$. By \Cref{prop:gamma_ordering}, only $\gamma$ affects quality rankings. Full proof in \Cref{app:proofs:regularization}.
\end{proofsketch}

The theorem has direct implications for when quality can be recovered. When the reviewer pool is ideologically neutral, the anchor vanishes, and BMF recovers quality:

\begin{corollary}[restate=corQualityRecovery,name=Quality Recovery under Neutrality]
\label{cor:quality_recovery}
Under the conditions of \Cref{thm:anchor_l2}, if $\mu_u = 0$ then BMF recovers the correct quality ranking for every configuration $(q, v)$. Proof in \Cref{app:proofs:regularization}.
\end{corollary}

The converse, whether $\mu_u \neq 0$ always breaks recovery, requires additional arguments and is established in \Cref{sec:tightness}. Missing the `objective' ranking is also not a noiseless artifact: it persists as ratings become precise.

\begin{corollary}[restate=corNoisyPersistence, name=Recovery Under Noise]
\label{cor:noisy_persistence}
For any reviewer pool with $\mu_u \neq 0$, the recovery failure is structural: as $\Delta \to 0$, the $L_2$ anchor satisfies $\gamma^*(\Delta) \to s\mu_u$ with $s \in (0,1)$, so for all sufficiently small $\Delta > 0$ BMF still misorders, with probability tending to one as $\Delta \to 0$, the configurations $(q, v)$ that it misorders at $\Delta = 0$. Proof in \Cref{app:proofs:regularization}.
\end{corollary}

When $\mu_u \neq 0$, it is natural to ask what BMF does recover. The answer is the average rating:

\begin{remark}[Average rating and aggregate welfare]
\label{rem:average_utility}
The average expected rating of note $j$ across all reviewers is $q_j + \mu_u \, v_j$, which equals the rating of a hypothetical reviewer with ideology $\mu_u$. Recovering objective quality (anchoring at $\gamma = 0$) and maximizing average reviewer satisfaction (anchoring at $\gamma = \mu_u$) therefore coincide only when $\mu_u = 0$.
\end{remark}

When $\mu_u \neq 0$, BMF's recovered ranking shifts away from objective quality and toward the ranking that maximizes average reviewer satisfaction, exactly to the latter in the $\kappa \to 0$ limit (\Cref{cor:normative_choice}). Whether this is desirable depends on whether the average reviewer is representative of the population whose welfare we care about, a normative question. The regularization weights determine where on this spectrum BMF lands:

\begin{corollary}[restate=corNormativeChoice, name=Regularization Encodes Design Choices]
\label{cor:normative_choice}
As $\kappa \to \infty$ (strong quality regularization relative to user regularization), $\gamma^* \to 0$: the \emph{ordering} of estimated qualities approaches the true quality ordering, though the absolute quality estimates are shrunk toward zero and may be uninformative. As $\kappa \to 0$ (strong user regularization), $\gamma^* \to \mu_u$: by \Cref{rem:average_utility}, the ordering matches the perspective of the average reviewer, maximizing aggregate reviewer satisfaction. Proof in \Cref{app:proofs:regularization}.
\end{corollary}

The two poles of \Cref{cor:normative_choice} are not equally robust to relaxing \Cref{thm:anchor_l2}'s moment conditions: the welfare pole at $\mu_u$ survives nonzero $\mathrm{Cov}(q,v)$, while the recovery pole at $0$ shifts: with $\mathrm{Cov}(q,v) \neq 0$ the agent-centric balance gives $\gamma^* = \bigl(t^*\mu_u - A\,\mathrm{Cov}(q,v)/\sigma_v^2\bigr)/(t^* + A)$ in the notation of \Cref{lem:unique_reduced}, whose limit is $\mu_u$ as $\kappa \to 0$ and $-\mathrm{Cov}(q,v)/\sigma_v^2$ as $\kappa \to \infty$. \Cref{fig:gamma_robustness_lambda_mu} confirms the $\gamma^* \propto \mu_u$ relationship is robust across regularization strengths. Either way, the choice of $\kappa$ via regularization is a design decision about whose preferences define quality.

\section{Non-Identifiability: No Mechanism on Ratings Alone}
\label{sec:tightness}

The recovery condition established in \Cref{sec:regularization} might appear specific to $L_2$ regularization. We now show that the gauge symmetry of \Cref{sec:gauge} lives in the rating data itself, so any mechanism on ratings alone faces the same fundamental ambiguity.

\begin{theorem}[Non-Identifiability of Quality]
\label{thm:quality-nonidentifiable}
In the ground truth model, for any two notes $j_1, j_2$ with $q_{j_1} > q_{j_2}$ and $v_{j_1} \neq v_{j_2}$, there exists a gauge-equivalent parameterization in which $q_{j_1}' < q_{j_2}'$. The quality ordering is reversed while all ratings remain unchanged.
\end{theorem}
\begin{proof}
Let $\delta = v_{j_2} - v_{j_1} \neq 0$. Choose $\gamma$ with $\gamma \delta > q_{j_1} - q_{j_2}$. By \Cref{lem:gauge_invariance}: $q_{j_2}' - q_{j_1}' = (q_{j_2} - q_{j_1}) + \gamma \delta > 0$.
\end{proof}
 
\begin{corollary}[restate=corNoMechanism, name=No mechanism can recover quality from ratings alone]
\label{cor:no_mechanism}
Let $M: \R^{n \times m} \to \R^m$ be any mechanism that maps a rating matrix to quality estimates. There exist ground-truth parameterizations $(u, q, v)$ and $(u', q', v')$ that produce identical rating matrices but with $q_{j_1} > q_{j_2}$ in one parameterization and $q'_{j_1} < q'_{j_2}$ in the other. Since $M$ must produce the same output for both, it cannot recover the correct quality ordering in both cases. Proof in \Cref{app:proofs:tightness}.
\end{corollary}
 
The ambiguity is not an artifact of BMF's particular algorithm but an inherent property of the rating data under the ground truth model. Without additional information about which reviewer ideology is truly neutral, the quality ordering is inherently ambiguous. Together with \Cref{cor:quality_recovery}, this gives the full characterization:

\begin{corollary}[Universal Quality Recovery]
\label{cor:quality_recovery_full}
Under the conditions of \Cref{thm:anchor_l2}, BMF recovers the correct quality ranking for every configuration $(q, v)$ if and only if $\mu_u = 0$.
\end{corollary}

\begin{proof}
$(\Leftarrow)$ \Cref{cor:quality_recovery}. $(\Rightarrow)$ If $\mu_u \neq 0$ then $\gamma^* \neq 0$. Fix $v$ with $\mu_v = 0$; since $\gamma^*$ depends on the configuration only through $\sigma_v^2$, this fixes $\gamma^* \neq 0$. Take two notes with $v_{j_1} \neq v_{j_2}$, labelled so that $\gamma^*(v_{j_2} - v_{j_1}) > 0$. Any $q$ with $\mu_q = 0$, $\mathrm{Cov}(q, v) = 0$, and $0 < q_{j_1} - q_{j_2} < \gamma^*(v_{j_2} - v_{j_1})$ then gives $\hat{q}_{j_2} > \hat{q}_{j_1}$, reversing the true order.
\end{proof}

\begin{figure}[t]
  \centering
  \begin{minipage}[c]{0.49\linewidth}
    \centering
    \includegraphics[width=\linewidth]{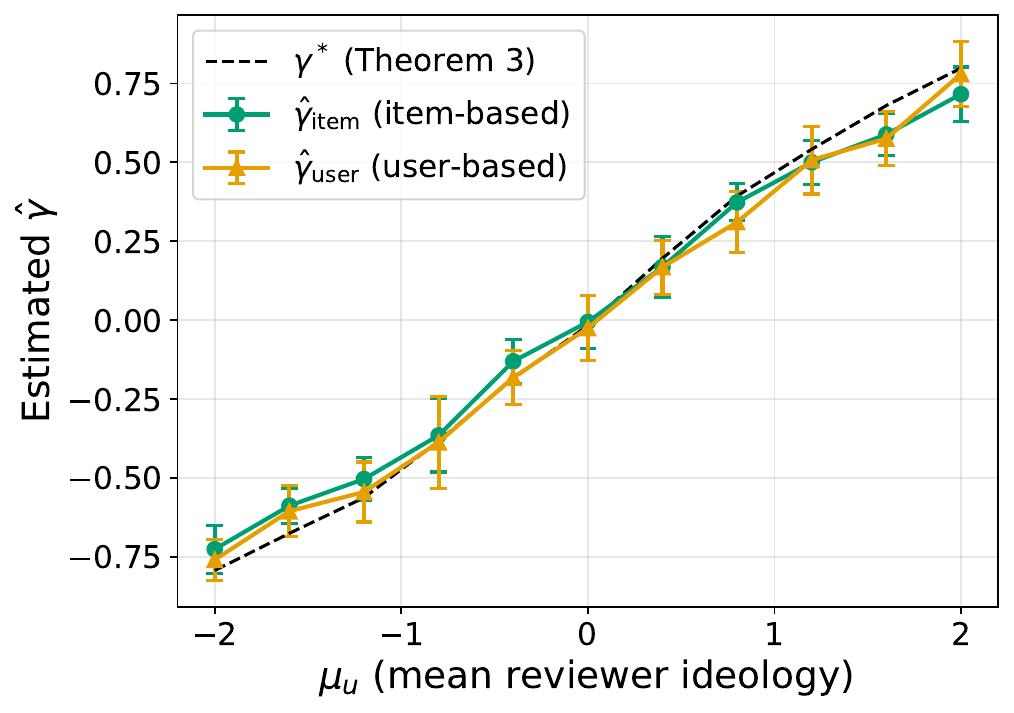}
    \caption{Estimated anchor $\gamma$ vs. mean reviewer ideology $\mu_u$ ($\kappa=1$). Item-based ($\hat{\gamma}_{\text{item}}$, circles) and user-based ($\hat{\gamma}_{\text{user}}$, triangles) estimators (\Cref{app:experiments}) both track $\gamma^*$ of \Cref{thm:anchor_l2} (dashed; the root of \eqref{eq:psi} at each replication's empirical moments, averaged within each $\mu_u$), which follows $\mu_u/2$ near zero and bends below it at the extremes. Error bars: $\pm 1$ SD over 10 replications.}
    \label{fig:gamma_sweep}
  \end{minipage}
  \hfill
  \begin{minipage}[c]{0.49\linewidth}
    \centering
    \includegraphics[width=\linewidth]{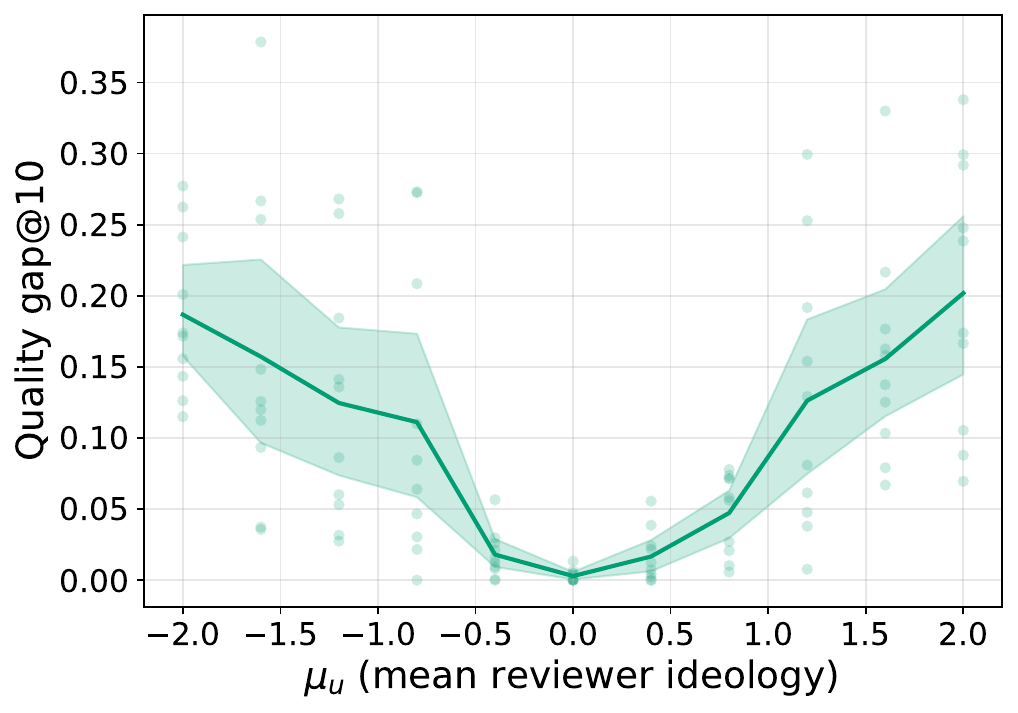}
    \caption{Quality gap@10 (top-10 quality vs.\ oracle, \Cref{app:experiments}) versus $\mu_u$. Near-zero at $\mu_u = 0$; grows monotonically with $|\mu_u|$. 10 replications ($\pm 1$ SE shaded).}
    \label{fig:quality_gap}
  \end{minipage}
\end{figure}

\section{Empirical Validation}
\label{sec:experiments}

We validate the theoretical predictions using simulations of the ground truth model with $n = 101$, $m = 100$, $d = 1$. Reviewer ideologies are drawn $u_i \sim \mathcal{N}(\mu_u, 1)$, notes from $q_j, v_j \sim \mathcal{N}(0, 1)$ independently, and ratings as $r_{ij} = q_j + u_i v_j + \varepsilon_{ij}$ with $\varepsilon_{ij} \sim \mathcal{N}(0, \Delta)$, $\Delta = 0.25$ ($\sigma = 0.5$). BMF is fitted with $L_2$ regularization using the Adam optimizer \citep{kingmaAdamMethodStochastic2017}. The anchor $\gamma$ is estimated by two complementary regressions: item side ($\hat{q}_j$ on $v_j$) and user side ($u_i$ on $\hat{u}_i$), which agree under correct specification. Full experimental details are in \Cref{app:experiments:gamma_estimation}.
 
\emph{$\gamma$ tracks $\mu_u$.} We vary $\mu_u \in [-2, 2]$ with $\lambda_u = \lambda_q = 0.01$ ($\kappa = 1$). In \Cref{fig:gamma_sweep} both estimators track $\gamma^*$ of \Cref{thm:anchor_l2}, the root of \eqref{eq:psi}, which follows $\mu_u/2$ near zero and bends below it at the extremes; at $\mu_u = 2$ the prediction is $0.80$ against $0.72$ (item-based) and $0.78$ (user-based). Individual estimates scatter around the line at these pool sizes, a finite-sample effect quantified in \Cref{app:experiments:robustness}. The relationship is robust across regularization strengths, ratio $\kappa$, and observation noise levels (\Cref{app:experiments:robustness}).
 
\emph{Quality recovery degrades with $|\mu_u|$.} \Cref{fig:quality_gap} shows that when $\mu_u = 0$, BMF achieves near-optimal quality recovery. As $|\mu_u|$ increases, quality is increasingly influenced by ideological alignment: notes leaning in the direction of the average reviewer are overrated.
 
\emph{Misspecified utilities.} We also generate ratings from three departures from the bilinear form: a Euclidean utility $-(u_i - v_j)^2$, a higher-order interaction $u_i v_j + \varphi\,u_i^2 v_j^2$, and an unmodeled confounder $u_i v_j + w_i z_j$. The anchor tracks $\mu_u$ under all three. Quality recovery at $\mu_u = 0$ survives the confounder and fails under the Euclidean utility and the higher-order interaction, which the theory allows since \Cref{cor:quality_recovery} is proved for the bilinear model. These are preliminary checks on three specific departures. Details in \Cref{app:experiments:misspecification}.

\emph{Binary ratings.} Community Notes uses categorical (helpful/not helpful) ratings. We test robustness by passing scores through a logistic link and sampling Bernoulli ratings. The user-based estimate tracks the curve of \Cref{thm:anchor_l2} ($0.77$ against $0.79$ at $\mu_u = 2$), while the item-based estimate compresses at large $|\mu_u|$ ($0.57$), likely due to saturation of the logistic link. This indicates that the anchoring to the population mean persists in the categorical setting used by Community Notes. Additional details in \Cref{app:experiments:binary}.

\section{Discussion and Conclusion}
\label{sec:discussion}
 
Under BMF, the decomposition of ratings into ``quality'' and ``ideology'' is not unique: shifting all reviewer ideologies by a constant while adjusting quality scores leaves every rating unchanged. BMF's regularization resolves this ambiguity by anchoring at a constant times the mean reviewer ideology (\Cref{thm:anchor_l2}), recovering true quality if and only if that mean is zero. Like the Condorcet Jury Theorem, our result identifies conditions for when a collective judgment procedure recovers truth. This limitation is not an $L_2$ artifact: no mechanism based on ratings alone can do better (\Cref{cor:no_mechanism}). The rest of this section draws out what these results imply for platform design and the broader question of what Community Notes is actually optimizing.
 
\emph{Indeterminacy and subjective vs.\ objective interpretations.} The gauge symmetries we identify reveal a core indeterminacy: ratings are consistent with the platform recovering an objective ground truth \emph{or} aggregating subjective preferences, depending solely on which ideology is declared neutral. Both interpretations map onto the same mathematical structure, differing only in the anchor. Our model assumes a single ground truth and asks whether BMF recovers it, but the indeterminacy suggests a more general framing in which there is no uniquely privileged vantage point. One could instead model quality as a family of equally valid standards, each anchored to a different perspective, with the platform's problem recast as choosing among them according to an explicit social welfare criterion \citep{arrowSocialChoiceIndividual2012, boutilierOptimalSocialChoice2015}. Nonetheless, our analysis reveals an asymmetry between these readings: the welfare anchor at $\mu_u$ is robust to moment conditions on $(q, v)$, while the truth-recovery anchor at $0$ requires them. In this sense BMF natively performs preference aggregation, with truth recovery available as a special case. %

\emph{Implications for Community Notes.} A skewed reviewer pool (likely, given self-selection \citep{allenBirdsFeatherDont2022}) means quality estimates inherit that skew: notes aligned with the majority ideology will rank higher not because they are better, but because BMF's anchor drifts toward the mean reviewer. Empirical assessments of ``helpfulness'' \citep{wojcikBirdwatchCrowdWisdom2022} carry the same confound, which follows from using ratings alone (\Cref{cor:no_mechanism}). The platform therefore faces a normative choice: whose perspective should define quality? External information could help where ratings alone cannot: expert-labeled ground truth examples fix the anchor to the expert's perspective \citep{zhaoInsightsComparativeStudy2023}; requiring high approval from both sides of the spectrum \citep{ovadyaBridgingSystemsOpen2023} can be read in the model as requiring $r_{ij} = q_j + u_i v_j$ to be high for reviewers with $u_i < 0$ and $u_i > 0$ alike, which holds only for notes with high $q_j$ and small $|v_j|$, whose estimate $q_j + \gamma^* v_j$ then barely depends on the anchor; and note-level markers of quality, such as whether a note cites sources, pin $q_j$ directly, which no gauge shift can absorb.

\emph{Limitations and future work.} We assume $d = 1$, matching Community Notes but potentially missing higher-dimensional disagreement. The non-identifiability extends to any $d$ (\Cref{thm:quality-nonidentifiable}), while characterizing the anchor there needs a separate analysis, since the balance equations become matrix equations. We focus on the correctly specified setting; a general treatment of misspecification remains open. We also analyze the core factorization on its own; a deployment adds missing ratings, weighting, and display thresholds on top of it, and filtering which reviewers count moves $\mu_u$ and with it the anchor. We analyze a static snapshot; dynamic effects including reviewer entry and exit and feedback loops \citep{akpinarLongtermDynamicsFairness2022} are left to future work. Developing adversarially robust BMF variants \citep{schoenebeckInformationElicitationRowdy2021}, using richer data to better disentangle quality from ideology, validating on real Community Notes data, and fully characterizing strategic incentives under BMF are important next steps. More broadly, our framework applies to any system learning latent representations to aggregate opinions: making normative and design choices explicit is a prerequisite for accountable algorithmic governance.

\begin{ack}
We thank Maggie Makar for helpful suggestions on the theoretical approach.
This work was supported by the National Science Foundation under awards 2313137 and 2208662. The authors declare no competing interests.
\end{ack}

\bibliographystyle{plainnat}
\bibliography{ornithologists}

\newpage
\appendix

\section{Notation Table}
\label{app:notation}
 
For convenience, we collect the paper-specific shorthand used across sections. Symbols defined at their only meaningful use site (e.g.,\ $\hat{r}_{ij}, \varepsilon_{ij}, \bot$), and standard mathematical notation, are not listed here. The full BMF and ground-truth parameters appear in \Cref{def:bmf,def:ground_truth}; the gauge family in \Cref{thm:gauge_exhaustive}.
 
\begin{center}
\renewcommand{\arraystretch}{1.15}
\begin{tabular}{@{}l@{\hspace{1.75em}}>{\raggedright\arraybackslash}p{0.62\linewidth}@{}}
\toprule
\textbf{Symbol} & \textbf{Meaning} \\
\midrule
\multicolumn{2}{@{}l}{\emph{Empirical statistics of the reviewer/note pool}} \\
$\mu_u, \mu_v$ & Mean reviewer ideology and mean note ideology, $\mu_u = \tfrac{1}{n}\sum_i u_i$, etc. \\
$\sigma_u^2, \sigma_v^2$ & Empirical variances of $u_i$ and $v_j$. \\
$\rho = m/n$ & Note-to-reviewer ratio. \\
\midrule
\multicolumn{2}{@{}l}{\emph{Loss decomposition (\Cref{def:bmf})}} \\
$R(\hat\theta)$ & Squared-error risk; the total loss is $\mathcal{L} = R + \Lambda$. \\
$\Lambda_{L_2}(\hat\theta)$ & $L_2$ regularizer with per-block weights $\lambda_u, \lambda_q, \lambda_v, \lambda_b, \lambda_c$ (\Cref{sec:regularization}). \\
$\lambda > 0$ & Overall regularization strength (common factor on all five weights, \Cref{prop:minmin_convergence}). \\
$\kappa = \lambda_q / \lambda_u$ & Quality-to-user weight ratio. \\
$\Delta$ & Noise variance ($\E[\varepsilon_{ij}^2] = \Delta$); noiseless setting is $\Delta = 0$. \\
\midrule
\multicolumn{2}{@{}l}{\emph{Gauge parameters (\Cref{thm:gauge_exhaustive}, \Cref{eq:full_gauge})}} \\
$\gamma$ & Agent-centric shift; the only ranking-relevant gauge parameter (\Cref{prop:gamma_ordering}). \\
$\eta$ & Item-centric shift. \\
$\tau$ & Rank-1 rescaling factor; $\tau > 0$ by the convention of \Cref{def:minmin}. \\
$\alpha, \beta$ & Additive offsets absorbed into intercepts and quality. \\
$\mathcal{G}(\theta_{\mathrm{true}})$ & Gauge orbit: the parameterizations predicting exactly as the true model, which minimize the risk for every $\Delta$ and achieve zero risk when $\Delta = 0$. \\
\midrule
\multicolumn{2}{@{}l}{\emph{Solution-related notation}} \\
$\theta^*_\Lambda$ & Min-min solution: the element of the gauge orbit minimizing $\Lambda$ (\Cref{def:minmin}). \\
$x^*$ & For any quantity $x$ derived from $\theta$, $x^*$ denotes its value at $\theta^*_\Lambda$. \\
$\gamma^* = s\,\mu_u$, $s \in (0,1)$ & Anchor selected by $L_2$ regularization (\Cref{thm:anchor_l2}). \\
\midrule
\multicolumn{2}{@{}l}{\emph{Reduced system (\Cref{lem:unique_reduced,lem:jacobian_nonsingular,lem:noise_finite})}} \\
$t = \tau^2$ & Squared scale parameter, the variable of the reduced scalar equation. \\
$A = \kappa\rho\sigma_v^2$ & Damping constant in $\gamma(t) = \mu_u t/(t + A)$. \\
$K = (\lambda_v/\lambda_u)\rho\sigma_v^2$ & Right side of the reduced scalar equation \eqref{eq:psi}. \\
$\psi$ & Left side of \eqref{eq:psi}; strictly increasing from $0$ to $\infty$. \\
$w$ & Moments of the base point, $(\mu_u, \sigma_u^2, \sigma_v^2, \mathrm{Cov}(q,v))$. \\
$\zeta$ & Deviation of the fitted moments from the true ones. \\
\bottomrule
\end{tabular}
\end{center}
 
\paragraph{Note on overloaded $\sigma$.}
In the experimental appendices, $\sigma$ denotes the observation-noise standard deviation (so $\Delta = \sigma^2$), while $\sigma(\,\cdot\,)$ denotes the logistic link in the binary-rating extension. The two uses are distinguished by the presence of an argument.

\newpage
\section{Additional Proofs}
\label{app:proofs}

\subsection{Gauge Symmetries (\Cref{sec:gauge})}
\label{app:proofs:gauge}

\thmGaugeExhaustive*
\begin{proof}
We prove the result for $d = 1$; the general case follows componentwise.

Consider any gauge transformation, assumed defined on all of parameter space and twice differentiable, satisfying
\begin{equation}\label{eq:gauge_constraint_app}
\hat{c}' + \hat{b}_i' + \hat{q}_j' + \hat{u}_i' \hat{v}_j' = \hat{c} + \hat{b}_i + \hat{q}_j + \hat{u}_i \hat{v}_j
\end{equation}
for all $i \in [n]$ and $j \in [m]$.

\textbf{Step 1: Affine structure.} Since the right-hand side is linear in $\hat{v}_j$, taking $\partial^2 / \partial \hat{v}_j^2$ of the left-hand side yields $\hat{u}_i' \frac{\partial^2 \hat{v}_j'}{\partial \hat{v}_j^2} + 2 \frac{\partial \hat{u}_i'}{\partial \hat{v}_j} \frac{\partial \hat{v}_j'}{\partial \hat{v}_j} = 0$. For this to hold for all possible values of $\hat{u}_i'$, we need $\frac{\partial^2 \hat{v}_j'}{\partial \hat{v}_j^2} = 0$, so $\hat{v}_j' = A_j \hat{v}_j + B_j$. By symmetry, $\hat{u}_i' = C_i \hat{u}_i + D_i$.

\textbf{Step 2: Universal coefficients.} Substituting the affine forms into \eqref{eq:gauge_constraint_app} and expanding:
\[\hat{c}' + \hat{b}_i' + \hat{q}_j' + C_i A_j \hat{u}_i \hat{v}_j + C_i B_j \hat{u}_i + D_i A_j \hat{v}_j + D_i B_j = \hat{c} + \hat{b}_i + \hat{q}_j + \hat{u}_i \hat{v}_j.\]
Taking $\frac{\partial}{\partial \hat{u}_i}$: $\frac{\partial \hat{b}_i'}{\partial \hat{u}_i} + C_i A_j \hat{v}_j + C_i B_j = \hat{v}_j$. For this to hold for all $\hat{v}_j$, we need $C_i A_j = 1$ for all $i, j$. Considering two items: $C_i A_{j_1} = 1 = C_i A_{j_2}$ implies $A_{j_1} = A_{j_2}$. Thus $A_j = A$ and $C_i = 1/A := C$ are universal. By the symmetric argument (differentiating with respect to $\hat{v}_j$), $B_j = B$ and $D_i = D$ are also universal.

\textbf{Step 3: Parametrization.} With $\hat{u}_i' = C \hat{u}_i + D$ and $\hat{v}_j' = \frac{1}{C}\hat{v}_j + B$, set $\tau = C$, $\gamma = -D/C$, $\eta = -\tau B$. The constant $DB = \gamma\eta$ from the product $\hat{u}_i'\hat{v}_j'$ goes into $\hat{c}'$, and the remaining terms into $\alpha, \beta$, which yields the five-parameter family in \eqref{eq:full_gauge}. The total degrees of freedom are $3 + 2d = 5$ for $d = 1$.
\end{proof}

\propGammaOrdering*
\begin{proof}
By the gauge invariance property, any gauge transformation satisfies $\hat{c} + \hat{b}_i + \hat{q}_j + \hat{u}_i \hat{v}_j = \hat{c}' + \hat{b}_i' + \hat{q}_j' + \hat{u}_i' \hat{v}_j'$ for all $i, j$.

Since $\hat{q}_j'$ depends only on $j$ (not on $i$), rearranging shows $\hat{q}_j'$ must have the form $\hat{q}_j' = \hat{q}_j + \langle \omega, \hat{v}_j \rangle + \beta$ for some $\omega \in \mathbb{R}^d$ and $\beta \in \mathbb{R}$ (by bilinearity of the interaction term and independence from agent index).

Now examine each basic gauge transformation:

\textit{Agent-centric transformation}: $\hat{q}_j \mapsto \hat{q}_j + \gamma \hat{v}_j$. This gives $\omega = \gamma$, which is nonzero for $\gamma \neq 0$.

\textit{Item-centric transformation}: $\hat{v}_j \mapsto \hat{v}_j - \eta$, $\hat{b}_i \mapsto \hat{b}_i + \hat{u}_i \eta$, $\hat{q}_j \mapsto \hat{q}_j$. Quality is unchanged, so $\omega = 0$.

\textit{Scaling}: $\hat{u}_i \mapsto \tau \hat{u}_i$, $\hat{v}_j \mapsto \hat{v}_j / \tau$. This preserves $\hat{u}_i \hat{v}_j$ and does not alter $\hat{q}_j$, so $\omega = 0$.

\textit{Global offsets}: $\hat{q}_j \mapsto \hat{q}_j + \beta$. This shifts all qualities by a constant, so $\omega = 0$.

Therefore, among the gauge transformations, only the agent-centric shift with $\gamma \neq 0$ produces $\omega \neq 0$ and can affect the relative ordering of qualities.
\end{proof}

\subsection{Quality Recovery (\Cref{sec:regularization})}
\label{app:proofs:regularization}

\lemBalanceLtwo*
\begin{proof}
For each gauge symmetry, the risk is invariant along the corresponding one-parameter family. Since the total objective $R + \lambda\Lambda$ must be stationary at the minimum, and the risk contribution is flat along the orbit, the derivative of $\Lambda$ alone must vanish in each gauge direction. 

For the agent-centric shift $\hat{u}_i \mapsto \hat{u}_i - \gamma$, $\hat{q}_j \mapsto \hat{q}_j + \gamma \hat{v}_j$: the $\gamma$-dependent regularization is $\Lambda(\gamma)$$ = \lambda_u \sum_i (\hat{u}_i - \gamma)^2 + \lambda_q \sum_j (\hat{q}_j + \gamma \hat{v}_j)^2 + \text{const}$. Setting $\frac{d\Lambda}{d\gamma}\big|_{\gamma=0} = 0$ yields (Agent-centric). 

For the scaling $\hat{u}_i \mapsto \tau \hat{u}_i$, $\hat{v}_j \mapsto \hat{v}_j / \tau$: the $\tau$-dependent part is $\Lambda(\tau)$$ = \lambda_u \sum_i (\tau \hat{u}_i)^2 + \lambda_v \sum_j (\hat{v}_j / \tau)^2$. Setting $\frac{d\Lambda}{d\tau}\big|_{\tau=1} = 0$ yields (Scale). 

For the intercept shift $\hat{b}_i \mapsto \hat{b}_i + \alpha$, $\hat{q}_j \mapsto \hat{q}_j + \beta$, $\hat{c} \mapsto \hat{c} - \alpha - \beta$: differentiating with respect to $\alpha$ and $\beta$ at zero yields (Intercept). 

For the item-centric shift $\hat{b}_i \mapsto \hat{b}_i + \hat{u}_i \eta$, $\hat{v}_j \mapsto \hat{v}_j - \eta$: differentiating at $\eta = 0$ yields (Item-centric).
\end{proof}

\thmAnchorLtwo*
\begin{proof}
We characterize the min-min solution under the ground truth model with $L_2$ regularization. The proof proceeds in three steps.

\subsection*{Step 1: Min-min characterization under the ground truth model}

Under the simplified ground truth model with $\mu_q = \mu_v = 0$ and $\mathrm{Cov}(q, v) = 0$, the min-min solution under $L_2$ regularization satisfies:
\begin{align}
\lambda_b\bigl[\tau n(\mu_u - \gamma)\eta + n\alpha\bigr] &= \lambda_q m\beta = -\lambda_c(\alpha + \beta) \tag{S1} \label{eq:S1} \\
\lambda_u\tau^2 n(\mu_u - \gamma) &= \lambda_q m(\gamma\sigma_v^2 - \beta\eta) \tag{S2} \label{eq:S2} \\
\lambda_u\tau^2 n[\sigma_u^2 + (\mu_u - \gamma)^2] &= \frac{\lambda_v m}{\tau^2}[\sigma_v^2 + \eta^2] \tag{S3} \label{eq:S3}
\end{align}
These follow from substituting the gauge transformation $\hat{u}_i = \tau(u_i - \gamma)$, $\hat{q}_j = q_j + \gamma v_j + \beta$, $\hat{v}_j = v_j/\tau - \eta/\tau$, $\hat{b}_i = u_i\eta + \alpha$, $\hat{c} = -\alpha - \beta - \gamma\eta$ into the balance equations of \Cref{lem:balance_l2}, then applying the moment assumptions.

\subsection*{Step 2: $\eta = \alpha = \beta = 0$ at the minimum}

By the law of iterated minimization: $\min_{\gamma,\eta,\tau} \Lambda(\gamma,\eta,\tau) = \min_\tau [\min_{\gamma,\eta} \Lambda(\gamma,\eta,\tau)]$.

For fixed $\tau$, the gauge orbit is affine in $(\gamma, \eta)$ and $\Lambda$ is strictly convex, giving a unique inner minimizer. From \eqref{eq:S1}, $\alpha$ and $\beta$ are linear in $\eta$: $\alpha(\eta) = k_1 \eta$ and $\beta(\eta) = k_2 \eta$ for constants $k_1, k_2$. Substituting, the $\eta$-dependent part of $\Lambda$ is quadratic:
\[\Lambda(\eta) = \Lambda(0) + \frac{2\lambda_v m \mu_v}{\tau^2}\eta + C\eta^2\]
where $C > 0$. Since $\mu_v = 0$ under the ground truth model, the linear term vanishes and $\eta^* = 0$ is the unique minimizer for every fixed $\tau$. Therefore $\eta = 0$ at the joint minimum. Since $b_i = c = 0$ in the ground truth, equation \eqref{eq:S1} then gives $\alpha = \beta = 0$.

\textit{Remark (weaker conditions).} The conclusion above extends to ground-truth models with $c + \mu_b = 0$ and $\mathrm{Cov}(b_i, u_i) = 0$, without requiring $c = b_i = 0$ pointwise. Under these conditions the inner minimizer becomes $(\alpha^*, \beta^*, \eta^*) = (-\mu_b, 0, 0)$ rather than $(0, 0, 0)$: the $\alpha$-shift absorbs the mean intercept into $\hat{c}$ (a residual gauge freedom under $c + \mu_b = 0$), while $\mathrm{Cov}(b_i, u_i) = 0$ removes the cross-term in the $\eta$ balance that arbitrary $b_i$ would otherwise contribute. The reduced system \eqref{eq:S2prime}--\eqref{eq:S3prime} for $(\gamma, \tau)$ contains no $b$- or $c$-dependent terms, so $\gamma$ remains a fraction of $\mu_u$.

\subsection*{Step 3: Solving the reduced system}

With $\eta = \alpha = \beta = 0$, the system reduces to:
\begin{align}
\lambda_u\tau^2 n(\mu_u - \gamma) &= \lambda_q m \gamma\sigma_v^2 \tag{S2'} \label{eq:S2prime} \\
\lambda_u\tau^2 n[\sigma_u^2 + (\mu_u - \gamma)^2] &= \frac{\lambda_v m\sigma_v^2}{\tau^2} \tag{S3'} \label{eq:S3prime}
\end{align}

\textbf{Solving for $\gamma$:} Rearranging \eqref{eq:S2prime}: $\lambda_u\tau^2 n\mu_u = \gamma(\lambda_u\tau^2 n + \lambda_q m\sigma_v^2)$, giving $\gamma = \mu_u / (1 + \kappa\rho\sigma_v^2/\tau^2)$.

\textbf{Solving for $\tau$:} Rearranging \eqref{eq:S3prime}: $\tau^4 = \lambda_v m\sigma_v^2 / (\lambda_u n[\sigma_u^2 + (\mu_u - \gamma)^2])$.

\textbf{Existence, uniqueness, and $s \in (0,1)$:} Substituting $\gamma(\tau^2)$ into \eqref{eq:S3prime} leaves a single equation in $\tau^2$ whose left side is continuous and strictly increasing from $0$ to $\infty$. It has exactly one root, which gives the unique fixed point $(\gamma^*, \tau^*)$ (\Cref{lem:unique_reduced}). Writing $\gamma^* = s\mu_u$: $s = 1/(1 + \kappa\rho\sigma_v^2/\tau^{*2})$. Since $\kappa, \rho, \sigma_v^2, \tau^{*2} > 0$, the denominator exceeds 1, so $s \in (0,1)$.

\textbf{Quality rankings:} By \Cref{prop:gamma_ordering}, only $\gamma$ affects the ordering of estimated qualities $\hat{q}_j = q_j + \gamma v_j$. The scale $\tau$ does not enter $\hat{q}_j$.
\end{proof}

\subsubsection*{Solving the reduced system}

Throughout, $t := \tau^2$ and
\[
A := \kappa\rho\sigma_v^2 = \frac{\lambda_q}{\lambda_u}\frac{m}{n}\sigma_v^2 > 0,
\qquad
K := \frac{\lambda_v}{\lambda_u}\frac{m}{n}\sigma_v^2 > 0 .
\]

\begin{lemma}[name=Uniqueness of the reduced system]
\label{lem:unique_reduced}
Assume $\sigma_u^2 > 0$ and $\mu_u > 0$. The system \eqref{eq:S2prime}--\eqref{eq:S3prime} has exactly one solution $(\gamma^*, \tau^*) \in (0, \mu_u) \times (0, \infty)$. It is given by $\gamma^* = \mu_u t^*/(t^* + A)$, where $t^*$ is the unique root of
\begin{equation}
\psi(t) := t^2\sigma_u^2 + \mu_u^2 A^2\Bigl(\frac{t}{t + A}\Bigr)^2 = K,
\label{eq:psi}
\end{equation}
and $\psi'(t) > 0$ for every $t > 0$. For $\mu_u < 0$ the solution is the sign reflection of the one for $|\mu_u|$; for $\mu_u = 0$ it is $\gamma^* = 0$ with $\tau^{*4} = K/\sigma_u^2$.
\end{lemma}

\begin{proof}
Equation \eqref{eq:S2prime} is linear in $\gamma$. Dividing by $\lambda_u n$ and using $\lambda_q m \sigma_v^2 = \lambda_u n A$ gives $t(\mu_u - \gamma) = A\gamma$, so
\begin{equation}
\gamma(t) = \frac{\mu_u t}{t + A}, \qquad \mu_u - \gamma(t) = \frac{\mu_u A}{t + A},
\label{eq:gamma_of_t}
\end{equation}
and $\gamma$ increases strictly from $\gamma(0) = 0$ to $\mu_u$. Equation \eqref{eq:S3prime} states $t^2[\sigma_u^2 + (\mu_u - \gamma)^2] = K$; substituting \eqref{eq:gamma_of_t} reduces the system to the scalar equation \eqref{eq:psi}.

Both terms of $\psi$ increase strictly on $(0,\infty)$, because $t \mapsto t^2$ and $t \mapsto t/(t+A)$ do, and
\[
\psi'(t) = 2t\sigma_u^2 + \frac{2\mu_u^2 A^3 t}{(t + A)^3} > 0 .
\]
Also $\psi(0) = 0$, and $\psi(t) \to \infty$ through its first term, which uses $\sigma_u^2 > 0$. A continuous strictly increasing function from $0$ to $\infty$ takes the value $K > 0$ exactly once, so \eqref{eq:psi} has a unique root $t^* \in (0,\infty)$ and $\gamma^* = \gamma(t^*) \in (0, \mu_u)$ is the unique solution of the system. Finally $s = \gamma^*/\mu_u = t^*/(t^* + A) \in (0,1)$, which is the form appearing in \eqref{eq:gamma_star}.

For $\mu_u < 0$, $\psi$ depends on $\mu_u$ only through $\mu_u^2$, so $t^*$ is unchanged while $\gamma(t)$ changes sign with $\mu_u$. For $\mu_u = 0$, \eqref{eq:gamma_of_t} gives $\gamma \equiv 0$ and \eqref{eq:psi} becomes $t^2\sigma_u^2 = K$.
\end{proof}

\begin{lemma}[name=Nonsingularity of the Jacobian]
\label{lem:jacobian_nonsingular}
Write the reduced system as $F(\gamma, t) = 0$ with
\[
F_1(\gamma, t) = \lambda_u n t(\mu_u - \gamma) - \lambda_q m \gamma\sigma_v^2,
\qquad
F_2(\gamma, t) = \lambda_u n t\bigl[\sigma_u^2 + (\mu_u - \gamma)^2\bigr] - \frac{\lambda_v m \sigma_v^2}{t} .
\]
At the solution of \Cref{lem:unique_reduced},
\[
\det J(\gamma^*, t^*) = -\frac{(\lambda_u n)^2 (t^* + A)}{t^*}\,\psi'(t^*) < 0 .
\]
The same conclusion holds in the coordinates $(\gamma, \tau)$, since $\tau \mapsto \tau^2$ is a diffeomorphism of $(0,\infty)$.
\end{lemma}

\begin{proof}
Using $\lambda_q m\sigma_v^2 = \lambda_u n A$, the partial derivatives are
\[
\begin{aligned}
\frac{\partial F_1}{\partial\gamma} &= -\lambda_u n(t + A),
&\qquad
\frac{\partial F_1}{\partial t} &= \lambda_u n(\mu_u - \gamma),
\\
\frac{\partial F_2}{\partial\gamma} &= -2\lambda_u n t(\mu_u - \gamma),
&
\frac{\partial F_2}{\partial t} &= \lambda_u n\bigl[\sigma_u^2 + (\mu_u - \gamma)^2\bigr] + \frac{\lambda_v m\sigma_v^2}{t^2} .
\end{aligned}
\]
Let $\gamma_1(t) = \mu_u t/(t + A)$ be the curve $F_1 = 0$ from \eqref{eq:gamma_of_t}, and set $g(t) := F_2(\gamma_1(t), t)$. Factoring out $\lambda_u n/t$,
\[
g(t) = \frac{\lambda_u n}{t}\Bigl(t^2\bigl[\sigma_u^2 + (\mu_u - \gamma_1(t))^2\bigr] - K\Bigr) = \frac{\lambda_u n}{t}\bigl(\psi(t) - K\bigr) .
\]
Since $F_1(\gamma_1(t), t) \equiv 0$, implicit differentiation gives $\gamma_1'(t) = -(\partial F_1/\partial t)/(\partial F_1/\partial\gamma)$, and the chain rule gives $g'(t) = (\partial F_2/\partial\gamma)\,\gamma_1'(t) + \partial F_2/\partial t$. Substituting $\gamma_1'$ and collecting over $\partial F_1/\partial\gamma$,
\[
\det J = \frac{\partial F_1}{\partial\gamma}\frac{\partial F_2}{\partial t} - \frac{\partial F_1}{\partial t}\frac{\partial F_2}{\partial\gamma} = \frac{\partial F_1}{\partial\gamma}\, g'(t) .
\]
At $t = t^*$ the identity $\psi(t^*) = K$ removes the derivative of the prefactor, leaving $g'(t^*) = \lambda_u n\,\psi'(t^*)/t^*$. With $\partial F_1/\partial\gamma = -\lambda_u n(t^* + A)$ and $\psi'(t^*) > 0$ from \Cref{lem:unique_reduced}, the determinant takes the stated form.
\end{proof}

\begin{remark}
\label{rem:transversal}
The negative sign states that the two curves $F_1 = 0$ and $F_2 = 0$ cross transversally at $(\gamma^*, t^*)$: the rows of $J$ are their normals, and parallel normals would force $\psi'(t^*) = 0$. Strict monotonicity of $\psi$ is the only input, and \Cref{lem:unique_reduced} establishes it before it establishes uniqueness.
\end{remark}

\begin{lemma}[name=Noise leaves the population anchor fixed]
\label{lem:noise_population}
Let $p_{ij}(\theta)$ denote the prediction of BMF at parameters $\theta$, and let $\mathcal{L}_0(\theta) = \sum_{i,j}\bigl(p_{ij}(\theta_{\mathrm{true}}) - p_{ij}(\theta)\bigr)^2$. For ratings $r_{ij} = p_{ij}(\theta_{\mathrm{true}}) + \varepsilon_{ij}$ with $\varepsilon_{ij}$ i.i.d.\ of mean zero and variance $\Delta$, drawn independently of $\theta$,
\[
\E\Bigl[\textstyle\sum_{i,j}\bigl(r_{ij} - p_{ij}(\theta)\bigr)^2\Bigr] = \mathcal{L}_0(\theta) + nm\Delta
\qquad\text{for every }\theta .
\]
Hence for every $\Delta \geq 0$ the minimizers of the expected risk are exactly $\mathcal{G}(\theta_{\mathrm{true}})$, the min-min problem is the one solved in \Cref{thm:anchor_l2}, and the anchor is $\gamma^* = s\mu_u$.
\end{lemma}

\begin{proof}
Expanding the square,
\[
\textstyle\sum_{i,j}\bigl(r_{ij} - p_{ij}(\theta)\bigr)^2 = \mathcal{L}_0(\theta) + 2\sum_{i,j}\varepsilon_{ij}\bigl(p_{ij}(\theta_{\mathrm{true}}) - p_{ij}(\theta)\bigr) + \sum_{i,j}\varepsilon_{ij}^2 .
\]
The middle term has mean zero, and the last has mean $nm\Delta$ and does not depend on $\theta$, so the expected loss differs from $\mathcal{L}_0$ by a constant. The risk is therefore minimized exactly when $p(\theta) = p(\theta_{\mathrm{true}})$, which is the definition of $\mathcal{G}(\theta_{\mathrm{true}})$. Noise raises the risk floor from $0$ to $\Delta$ per entry and leaves the minimizing set unchanged. \Cref{prop:minmin_convergence} then applies with the excess risk $R - nm\Delta$ in place of the risk. Neither the constraint set nor $\Lambda_{L_2}$ depends on $\Delta$, so Steps 1 to 3 above apply verbatim and return the same $\gamma^*$.
\end{proof}

\begin{remark}
\label{rem:gauge_flat_under_noise}
The empirical loss depends on $\theta$ only through the predictions $p(\theta)$, which are constant along gauge orbits (\Cref{lem:gauge_invariance}). For every realization of the noise it is therefore constant on each orbit, and its derivative in any gauge direction is zero. A minimizer of the regularized objective minimizes $\Lambda$ along its own orbit, so the balance equations \eqref{eq:balance} hold exactly, with no noise correction. Noise changes which orbit the fit lands on while leaving its position along that orbit fixed. \Cref{lem:noise_finite} bounds the displacement of the anchor that results.
\end{remark}

\begin{lemma}[name=Displacement of the anchor at finite sample size]
\label{lem:noise_finite}
Steps 1 to 3 use the base point only through the moments $w := (\mu_u, \sigma_u^2, \sigma_v^2, \mathrm{Cov}(q, v))$, so write the reduced system as $F(\gamma, t; w) = 0$, with the additional term $-\lambda_q m\,\mathrm{Cov}(q,v)$ in $F_1$ and $F$ otherwise as in \Cref{lem:jacobian_nonsingular}. Let $w_0 = (\mu_u, \sigma_u^2, \sigma_v^2, 0)$ denote the true moments, and suppose the fit lies on the gauge orbit of a base point whose moments $\tilde{w}$ satisfy $\|\tilde{w} - w_0\| \leq \zeta$. There are $\zeta_0 > 0$ and $c_J < \infty$, depending only on $w_0$ and the regularization weights, such that for $\zeta < \zeta_0$ the system $F(\gamma, t; \tilde{w}) = 0$ has a solution $(\gamma_\Delta, t_\Delta)$, unique in a neighborhood of $(\gamma^*, t^*)$, with
\[
|\gamma_\Delta - \gamma^*| + |t_\Delta - t^*| \leq c_J\,\zeta .
\]
\end{lemma}

\begin{proof}
$F$ is a polynomial in $(\gamma, w)$ and in $t$ and $t^{-1}$, hence continuously differentiable on $(0,\mu_u) \times (0,\infty) \times \R^4$, and $F(\gamma^*, t^*; w_0) = 0$ by \Cref{lem:unique_reduced}. Its Jacobian in $(\gamma, t)$ at that point is the matrix of \Cref{lem:jacobian_nonsingular}, invertible with $\|J^{-1}\|$ bounded through the closed form of $\det J$. The implicit function theorem gives a neighborhood on which the perturbed system has a unique root, with
\[
\begin{pmatrix}\gamma_\Delta - \gamma^*\\ t_\Delta - t^*\end{pmatrix}
= -J^{-1}\,\frac{\partial F}{\partial w}\Bigl|_{(\gamma^*, t^*, w_0)}(\tilde{w} - w_0) + o(\zeta),
\]
so $c_J = \|J^{-1}\|\,\|\partial F/\partial w\| + o(1)$ suffices.

Under a perturbed base point $\mu_v$ is no longer exactly zero, so Step 2 gives $\eta^*, \alpha^*, \beta^* = O(\zeta)$ in place of $0$. These enter \eqref{eq:S2}--\eqref{eq:S3} only through $\beta\eta$ and $\eta^2$, contributing $O(\zeta^2)$, below the order of the remainder.
\end{proof}

\corQualityRecovery*
\begin{proof}
$\mu_u = 0 \Rightarrow \gamma^* = 0$ by \eqref{eq:gamma_star}, so $\hat{q}_j = q_j + \beta$ and the ranking is preserved.
\end{proof}

\corNoisyPersistence*
\begin{proof}
By \Cref{lem:noise_population} the expected loss under noise differs from the noiseless loss by the constant $nm\Delta$, so the risk minimizing set, the reduced system \eqref{eq:S2prime}--\eqref{eq:S3prime}, and the anchor $\gamma^* = s\mu_u$ are the same for every $\Delta \geq 0$. For $\mu_u \neq 0$ this gives $\gamma^*(\Delta) = s\mu_u \neq 0$, so by \Cref{prop:gamma_ordering} BMF misorders every pair with $0 < q_{j_1} - q_{j_2} < \gamma^*(v_{j_2} - v_{j_1})$, and configurations with such pairs exist (\Cref{cor:quality_recovery_full}).

At finite sample size the fit leaves $\mathcal{G}(\theta_{\mathrm{true}})$ and lies on the orbit of a perturbed base point (\Cref{rem:gauge_flat_under_noise}). If its moments deviate by $\zeta$, \Cref{lem:noise_finite} gives $\gamma_\Delta = \gamma^* + O(\zeta)$, and the conclusion persists as long as $\zeta$ stays below the level at which $\gamma_\Delta$ could reach $0$. Sums of the form $\sum_{i,j}\varepsilon_{ij}z_{ij}$ with weights of bounded second moment concentrate at scale $\sqrt{nm\Delta}$, so against the $O(nm)$ scale of the deterministic terms $\zeta = O_p(\sqrt{\Delta/(nm)})$, and in particular $\zeta = O_p(\sqrt{\Delta})$ uniformly in $n$ and $m$.
\end{proof}

\corNormativeChoice*
\begin{proof}
From \eqref{eq:gamma_star}, $\gamma^* = \mu_u / (1 + \kappa\rho\sigma_v^2/\tau^{*2})$. As $\kappa \to \infty$ the denominator diverges, so $\gamma^* \to 0$ and the quality ordering is recovered. As $\kappa \to 0$ the denominator approaches $1$, so $\gamma^* \to \mu_u$; by \Cref{rem:average_utility} this matches the perspective of the average reviewer.
\end{proof}

\subsection{Non-Identifiability (\Cref{sec:tightness})}
\label{app:proofs:tightness}

\corNoMechanism*
\begin{proof}
Both parameterizations $(u, q, v)$ and $(u', q', v')$ yield identical rating matrices by \Cref{thm:quality-nonidentifiable}. Any mechanism $M$ operating on the rating matrix alone must return the same output for both. Since their quality orderings are reversed, $M$ cannot recover the correct ordering in both cases.
\end{proof}

\section{Running Example Details}
\label{app:running_example}

We illustrate the paper's main developments through a single scenario small enough that every
quantity can be verified by hand. Three reviewers with ideologies $u_1=-0.5$, $u_2=0$, $u_3=2$
evaluate two notes. Population statistics: $\mu_u = 0.5$, $\sigma_u^2 \approx 1.167$;
$\mu_v = 0$, $\sigma_v^2 = 1.0$, $\rho = m/n = 2/3$. The moment conditions of \Cref{thm:anchor_l2} hold approximately: $\mu_v = 0$ exactly, $\mu_q = 0.7$ only adds a common offset to the recovered qualities, and the quality-leaning covariance $\tfrac{1}{m}\sum_j q_j v_j = -0.1$ is small.

\subsection*{Ratings}

\begin{table}[h]
\centering
\small
\begin{tabular}{lcc}
\toprule
 & Note A ($q=0.8$, $v=-1$) & Note B ($q=0.6$, $v=1$) \\
\midrule
$u = -0.5$ & $1.30$ & $0.10$ \\
$u = 0\phantom{.0}$ & $0.80$ & $0.60$ \\
$u = 2\phantom{.0}$ & $-1.20$ & $2.60$ \\
\bottomrule
\end{tabular}
\caption{Running example ratings $r_{ij} = q_j + u_i v_j$ (simplified model, $c = b_i = 0$).
The neutral reviewer ($u=0$) rates by quality alone; all other rows reflect ideological alignment.
True quality ranking: $A > B$.}
\label{tab:ratings}
\end{table}

\subsection*{Gauge shift ($\gamma = 1$)}

Applying $\gamma = 1$ yields $q'_A = 0.8 + (1)(-1) = -0.20$, $q'_B = 0.6 + (1)(1) = 1.60$,
and shifted ideologies $u'_i = u_i - 1$: $u'_1 = -1.5$, $u'_2 = -1$, $u'_3 = 1$.
Every rating is preserved:

\begin{table}[h]
\centering
\small
\begin{tabular}{lcc}
\toprule
 & Note A ($q'=-0.20$, $v=-1$) & Note B ($q'=1.60$, $v=1$) \\
\midrule
$u' = -1.5$ & $1.30$ & $0.10$ \\
$u' = -1\phantom{.0}$ & $0.80$ & $0.60$ \\
$u' = 1\phantom{.0}$ & $-1.20$ & $2.60$ \\
\bottomrule
\end{tabular}
\caption{Gauge-shifted parameterization ($\gamma=1$). All entries are identical to
\Cref{tab:ratings}, but quality ranking is reversed: $q'_B > q'_A$.
Confirms \Cref{thm:quality-nonidentifiable}.}
\label{tab:ratings:shift}
\end{table}

\subsection*{BMF anchor and ranking flip ($\kappa=1$)}

Solving the balance equations with $\kappa=1$, $\rho=2/3$, $\sigma_v^2=1$,
$\sigma_u^2\approx 1.167$, $\mu_u=0.5$ gives $\gamma^*\approx 0.263$, $\tau^*\approx 0.859$.
Recovered qualities:
\[
  \hat{q}_A = 0.8 + 0.263\times(-1) = 0.537, \qquad
  \hat{q}_B = 0.6 + 0.263\times 1 = 0.863.
\]
The flip threshold is $(q_A-q_B)/(v_B-v_A) = 0.20/2 = 0.10$.
Since $\gamma^* = 0.263 > 0.10$, BMF ranks Note~B above Note~A with default regularization. (Minimizing $\Lambda_{L_2}$ over the full gauge orbit of this example, which accounts for the small covariance, gives $\gamma^* \approx 0.31$ and the same ranking.)

\newpage
\section{Experimental Details}
\label{app:experiments}
 
This appendix collects the implementation, hyperparameter, hardware, and evaluation-metric details for the simulations reported in \Cref{sec:experiments}.
 
\subsection{Data Generating Process (DGP)}
\label{app:experiments:dgp}
 
All experiments instantiate the simplified ground truth setting (\Cref{def:ground_truth} with $c = b_i = 0$, $d = 1$). The default DGP draws agent ideologies $u_i \stackrel{\text{i.i.d.}}{\sim} \mathcal{N}(\mu_u, 1)$ (with $\mu_u = 0$ unless varied) and item parameters $(q_j, v_j) \stackrel{\text{i.i.d.}}{\sim} (\mathcal{N}(0,1), \mathcal{N}(0,1))$ independently. Continuous ratings are generated as $r_{ij} = q_j + v_j u_i + \varepsilon_{ij}$ with $\varepsilon_{ij} \sim \mathcal{N}(0, \Delta)$ and $\Delta = 0.25$ ($\sigma = 0.5$). The default sample sizes are $n = 101$ agents and $m = 100$ items. Every experiment refers to this default setup and only describes parameters that differ.
 
\subsection{Implementation and Optimization}
\label{app:experiments:implementation}
 
We use our own \texttt{PyTorch} \citep{paszkePyTorchImperativeStyle2019} implementation of BMF (\Cref{def:bmf}), which makes it straightforward to swap in different regularizers, explicit gauge constraints, and link functions. Ideology factors are initialized Xavier-uniform, intercepts and qualities at zero, and all parameters are trained with the Adam optimizer \citep{kingmaAdamMethodStochastic2017} with learning rate $0.05$ ($0.01$ for binary ratings) for $500$ epochs of full-batch training, with early stopping on the training loss (patience $10$, tolerance $10^{-5}$). The training objective is mean squared error for continuous ratings and binary cross-entropy (log loss) for binary ratings. Regularization is $L_2$, with each parameter block penalized by the mean of its squares: weight $\lambda$ on $\hat{u}$ and $\hat{v}$, and $\kappa\lambda$ on $\hat{b}$, $\hat{q}$, and $\hat{c}$, with $\lambda = 0.01$ and $\kappa = 1$ by default. In the sum convention of \Cref{sec:regularization} this is $\lambda_u = \lambda m$, $\lambda_v = \lambda n$, $\lambda_q = \kappa\lambda n$, $\lambda_b = \kappa\lambda m$, so $A = \kappa\sigma_v^2$ and $K = \sigma_v^2$ in \Cref{lem:unique_reduced}, which for $n \approx m$ is the setting the figures compare against.
 
\subsection{Reproducibility and Compute}
\label{app:experiments:compute}
 
Experiments are run on a MacBook Pro with an Apple~M1~Pro chip. Each individual simulation at the default settings ($n=101$, $m=100$, $500$ epochs) takes approximately $10$ seconds. The codebase is parallelized over replications and configurations, so the full suite of runs reported in this paper (sweeps over $\mu_u$, regularization strength, the regularization ratio $\kappa$, the binary-rating variant, and $10$ replications per configuration) completes in a few hours on this machine. Code and instructions to reproduce all figures are available at \url{https://github.com/BibliotecaBabel/BirdwatchBMFAnalysis}.
 
\subsection{Estimating the Gauge Shift $\gamma$}
\label{app:experiments:gamma_estimation}
 
The agent-centric gauge transformation maps $\hat{u}_i \mapsto \hat{u}_i - \gamma$ and $\hat{q}_j \mapsto \hat{q}_j + \gamma \hat{v}_j$ while leaving predicted ratings unchanged. Given access to the true simulation parameters, we estimate $\gamma$ in two complementary ways:
\begin{description}
    \item[Item-based ($\hat{\gamma}_{\text{item}}$).] Fit the regression $\hat{q}_j = \beta_0 + \beta_1 v_j + \epsilon_j$ and take $\hat{\gamma}_{\text{item}} = \hat{\beta}_1$. This captures the systematic covariation between estimated quality and item ideology.
    \item[User-based ($\hat{\gamma}_{\text{user}}$).] Fit the regression $u_i = \beta_0 + \beta_1 \hat{u}_i + \epsilon_i$ and take $\hat{\gamma}_{\text{user}} = \hat{\beta}_0$. On the orbit $\hat{u}_i = \tau(u_i - \gamma)$, so $u_i = \gamma + \hat{u}_i/\tau$ and the intercept is $\gamma$ for every $\tau$.
\end{description}
Under correct specification both estimators converge to the same value, and our experiments report close agreement (within $\pm 1$ standard error; see \Cref{fig:gamma_sweep}).
 
\subsection{Evaluation Metrics}
\label{app:experiments:metrics}

For a mechanism $M$ producing estimated qualities $\hat{q}$, let $S(M)$ be the set of $\ell$ items selected as the top-$\ell$ by $\hat{q}_j$; we use $\ell = 10$.

\paragraph{Reference mechanisms (oracles).} We compare the BMF mechanism against an idealized benchmark that has access to ground-truth information unavailable to any real mechanism.
\begin{description}
    \item[Quality Oracle ($M^*_q$).] Selects the $\ell$ items with the highest true quality $q_j$. This is the best any mechanism can do on objective quality (OQ, defined below), and serves as the upper bound for quality-based selection.
\end{description}
This oracle is unachievable in practice since it requires knowledge of true qualities; it serves purely as a normalization baseline for the gap metric below.

\paragraph{Objective Quality gap.} Normalized shortfall in total true quality of selected items relative to the Quality Oracle $M^*_q$:
\[
\mathrm{OQ}_{\text{gap}}(M) = \frac{\mathrm{OQ}(M^*_q) - \mathrm{OQ}(M)}{\mathrm{OQ}(M^*_q)}, \qquad \mathrm{OQ}(M) = \sum_{j \in S(M)} q_j.
\]
This is the metric reported as \emph{Quality gap@10} in \Cref{fig:quality_gap}.

The metric is reported as the mean over $10$ replications, with error bars showing $\pm 1$ standard error.
 
\subsection{Robustness Checks}
\label{app:experiments:robustness}
 
We verify the robustness of the $\gamma$--$\mu_u$ relationship along four axes.

\paragraph{Variation across replications.} The pool sizes $n = 101$ and $m = 100$ match the sparse-data regime of a deployed system and keep the simulations away from large-$n$ asymptotics. Individual estimates $\hat{u}_i$ and $\hat{q}_j$ vary widely across replications at these sizes, and so does $\hat{\gamma}$. The spread follows from \Cref{lem:noise_finite}: the empirical moments of a finite draw deviate from their population values by $O_p(n^{-1/2})$, so the anchor fluctuates around $\gamma^*$ at the same order. The scatter therefore centers on $\gamma^*$. The drift toward $\mu_u$ is stable across replications.
 
\begin{figure}[h]
    \centering
    \includegraphics[width=\linewidth]{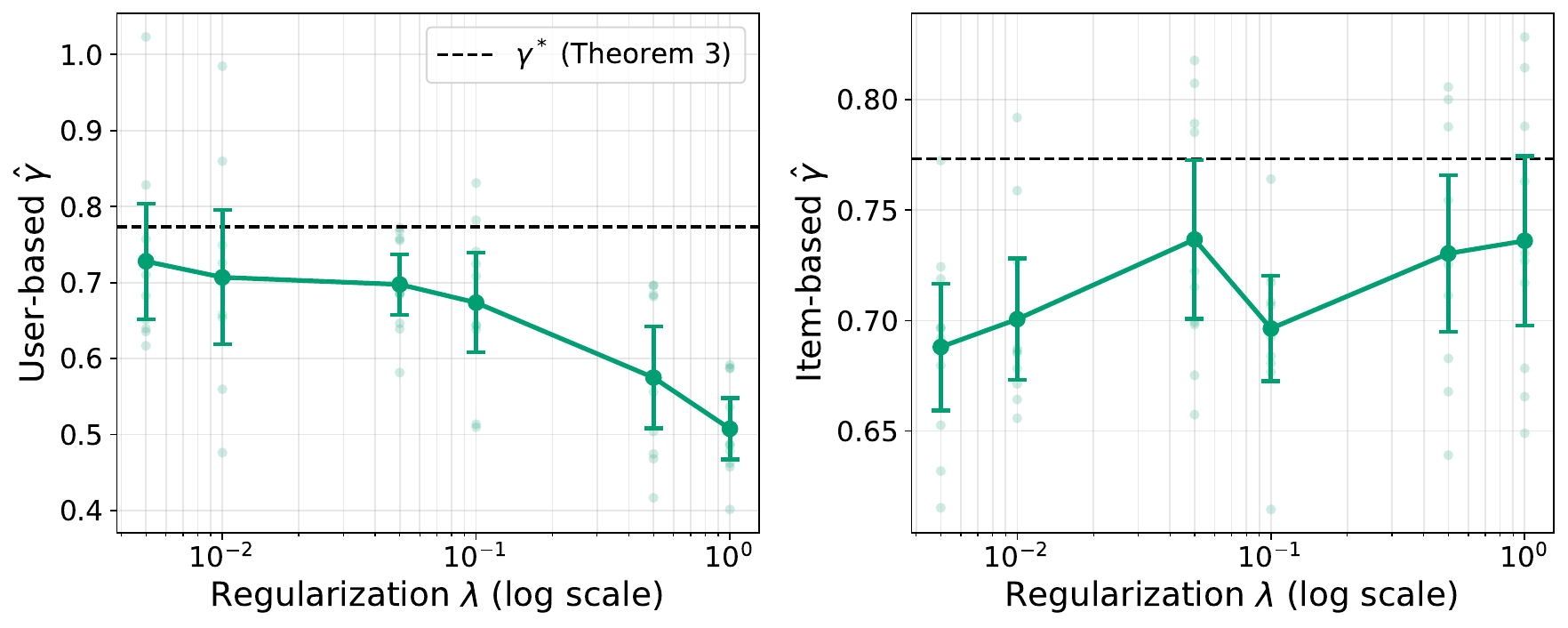}
    \caption{Estimated $\hat{\gamma}$ vs.\ regularization strength $\lambda$ ($\mu_u=2$, $\kappa=1$, $\sigma=0.1$). Dashed: $\gamma^* = 0.78$ from \eqref{eq:psi}. Estimates stay within about $0.1$ of it for $\lambda \leq 0.1$; the user-based estimate falls to $0.51$ at $\lambda = 1$, where the fit trades risk for norm (\Cref{fig:gamma_robustness_risk}).}
    \label{fig:gamma_robustness_lambda}
\end{figure}

\begin{figure}[h]
    \centering
    \includegraphics[width=0.8\linewidth]{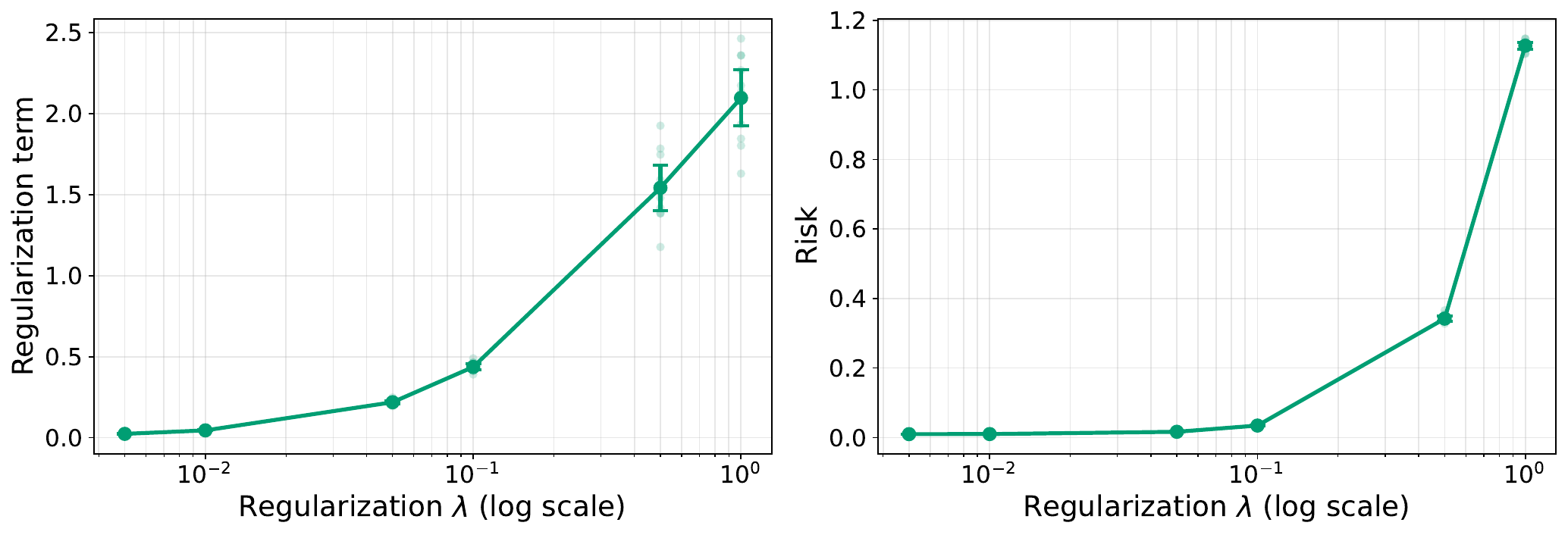}
    \caption{Regularization term (left) and training risk (right) vs.\ $\lambda$. Risk stays near its minimum for $\lambda \leq 0.1$; beyond that, the optimizer trades predictive accuracy for lower parameter norm, explaining the deviation in $\hat{\gamma}$ at high $\lambda$.}
    \label{fig:gamma_robustness_risk}
\end{figure}

\begin{figure}[h]
    \centering
    \includegraphics[width=\linewidth]{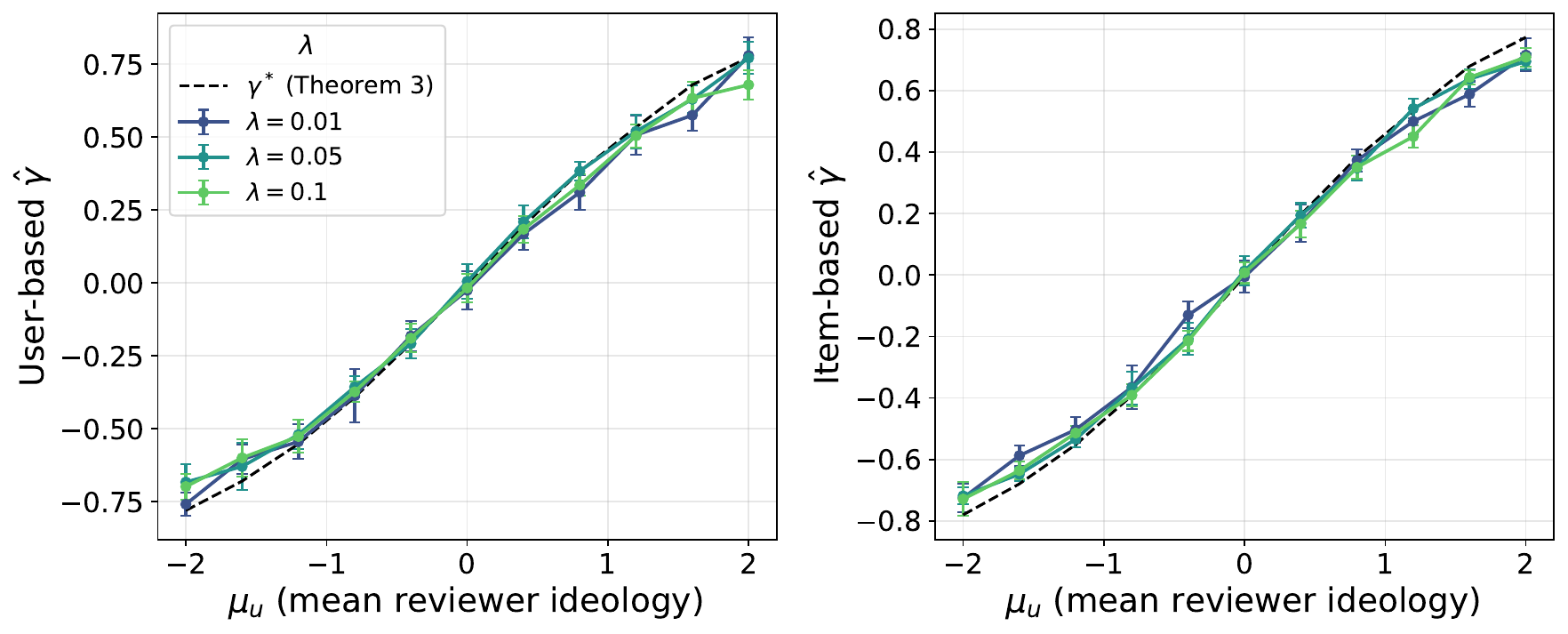}
    \caption{Estimated $\hat{\gamma}$ vs.\ $\mu_u$ for different regularization strengths $\lambda$ (L2, $\kappa=1$). The three curves coincide and track $\gamma^*$ of \Cref{thm:anchor_l2} (dashed), following $\mu_u/2$ near zero and bending below it at the extremes, which shows that the anchor is not sensitive to the choice of regularization strength.}
    \label{fig:gamma_robustness_lambda_mu}
\end{figure}

\paragraph{Regularization strength.} Fixing $\mu_u = 2$, $\kappa = 1$, and $\sigma = 0.1$, and varying $\lambda = \lambda_u = \lambda_q$ from $5 \times 10^{-3}$ to $10^{0}$: the estimates stay within about $0.1$ of $\gamma^* = 0.78$ for $\lambda \leq 0.1$, with a small dip at $\lambda = 0.005$ ($0.69$ item-based, $0.73$ user-based); the item-based estimate stays between $0.69$ and $0.74$ over the whole range, while the user-based estimate falls for $\lambda \geq 0.5$, to $0.51$ at $\lambda = 1$, as the optimizer trades predictive accuracy for lower parameter norms. \Cref{fig:gamma_robustness_lambda} shows the estimators; \Cref{fig:gamma_robustness_risk} shows the corresponding training risk and regularization term, confirming that risk begins to rise where $\hat{\gamma}_{\text{user}}$ deviates. \Cref{fig:gamma_robustness_lambda_mu} replicates the full $\gamma$--$\mu_u$ sweep across several $\lambda$ values; the three curves coincide.

\begin{figure}[h]
    \centering
    \includegraphics[width=\linewidth]{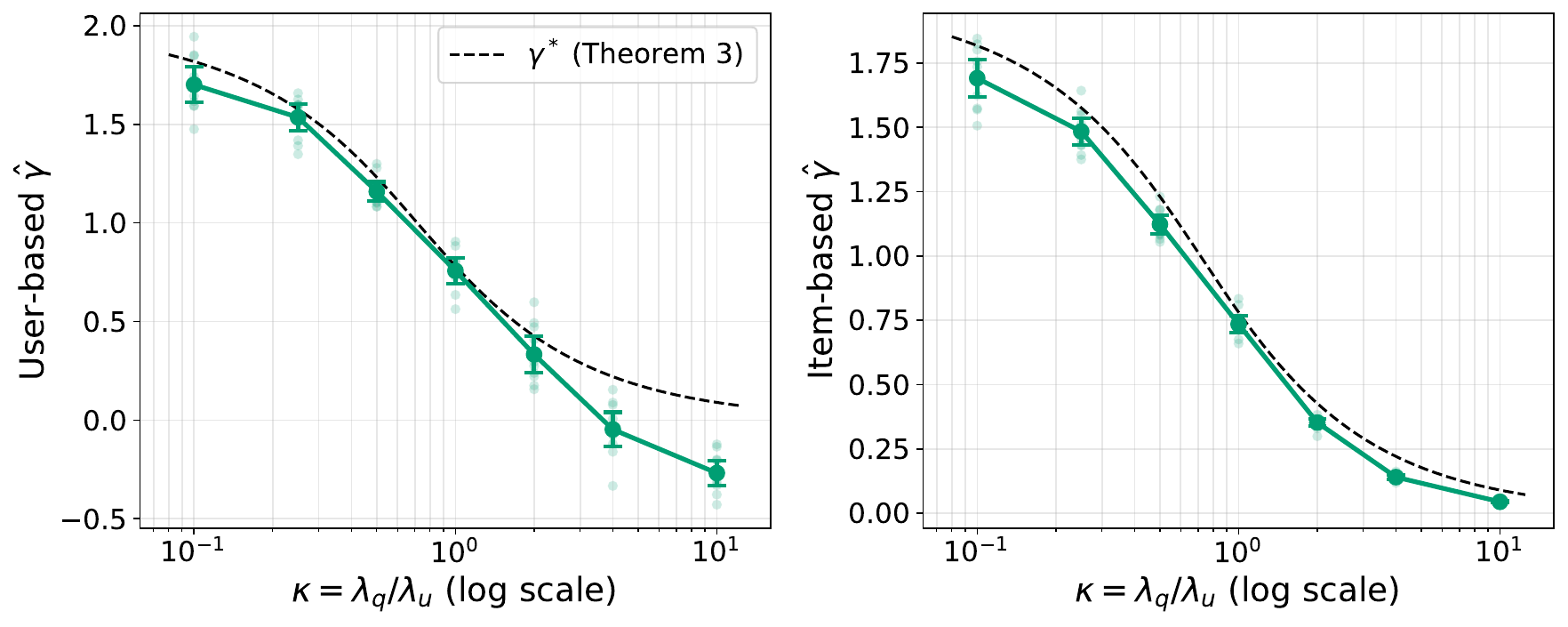}
    \caption{Estimated $\hat{\gamma}$ vs.\ regularization ratio $\kappa = \lambda_q/\lambda_u$ ($\mu_u=2$, $\lambda=0.1$, log scale). Dashed: $\gamma^*(\kappa)$ from \eqref{eq:psi}. The item-based estimate follows it across two decades of $\kappa$; the user-based estimate tracks it up to $\kappa = 2$ and undershoots beyond.}
    \label{fig:gamma_robustness_kappa}
\end{figure}

\paragraph{Regularization ratio $\kappa$.} Fixing $\mu_u = 2$ and $\lambda = 0.1$ and varying $\kappa$ from $0.1$ to $10$ (\Cref{fig:gamma_robustness_kappa}): the item-based estimates ($1.69$, $1.48$, $1.12$, $0.74$, $0.35$, $0.14$, $0.04$) follow $\gamma^*(\kappa)$ from \eqref{eq:psi} ($1.82$, $1.61$, $1.21$, $0.80$, $0.44$, $0.23$, $0.09$) and sit well below $\mu_u/(1+\kappa)$, the value at $\tau^{*2} = 1$; the user-based estimate tracks the curve up to $\kappa = 2$ and undershoots beyond it ($-0.05$ at $\kappa = 4$, $-0.27$ at $\kappa = 10$). As $\kappa$ grows (quality factors penalized more heavily relative to user factors), $\gamma$ decreases; as $\kappa$ shrinks, $\gamma$ increases toward $\mu_u$. This sweep uses $\lambda = 0.1$ because at $\lambda = 0.01$ the fit stalls near $0.89$ for $\kappa \leq 0.5$: along the gauge orbit the risk is flat, so the descent there is driven by the regularizer alone and is slow at small $\lambda$.

\begin{figure}[h]
    \centering
    \includegraphics[width=\linewidth]{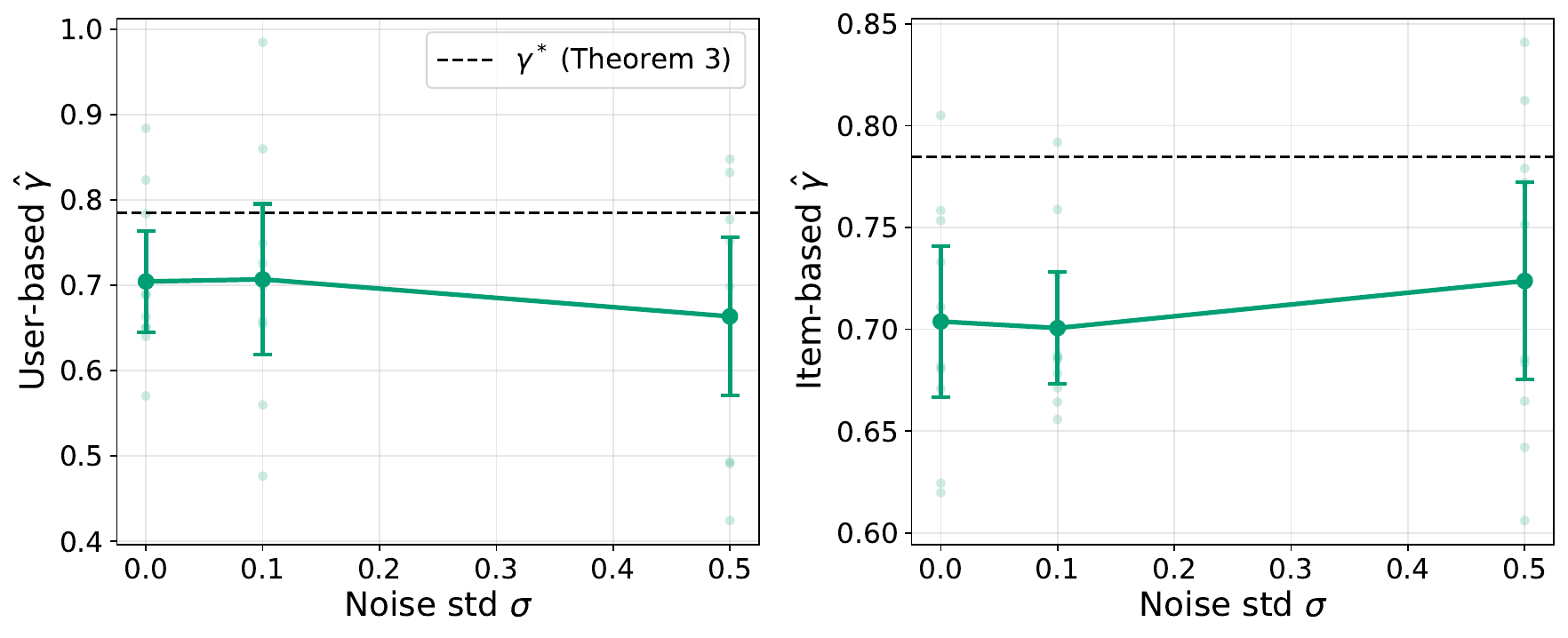}
    \caption{Estimated $\hat{\gamma}$ vs.\ observation noise $\sigma$ ($\mu_u=2$, L2, $\kappa=1$, $\lambda=0.01$). The estimators remain close to $\gamma^* = 0.78$ (dashed, from \eqref{eq:psi}) across the noise levels tested.}
    \label{fig:gamma_robustness_noise}
\end{figure}

\paragraph{Noise level.} We additionally vary observation noise $\sigma$ while holding all other parameters fixed ($\mu_u=2$, L2, $\kappa=1$, $\lambda=0.01$). Estimates remain close to $\gamma^* = 0.78$ across all noise levels tested (item-based $0.70$, $0.70$, $0.72$ and user-based $0.71$, $0.71$, $0.66$ at $\sigma = 0, 0.1, 0.5$), indicating that the anchor is robust to moderate measurement noise (\Cref{fig:gamma_robustness_noise}).

\subsection{Binary Ratings}
\label{app:experiments:binary}
 
\begin{figure}[h]
    \centering
    \includegraphics[width=0.6\linewidth]{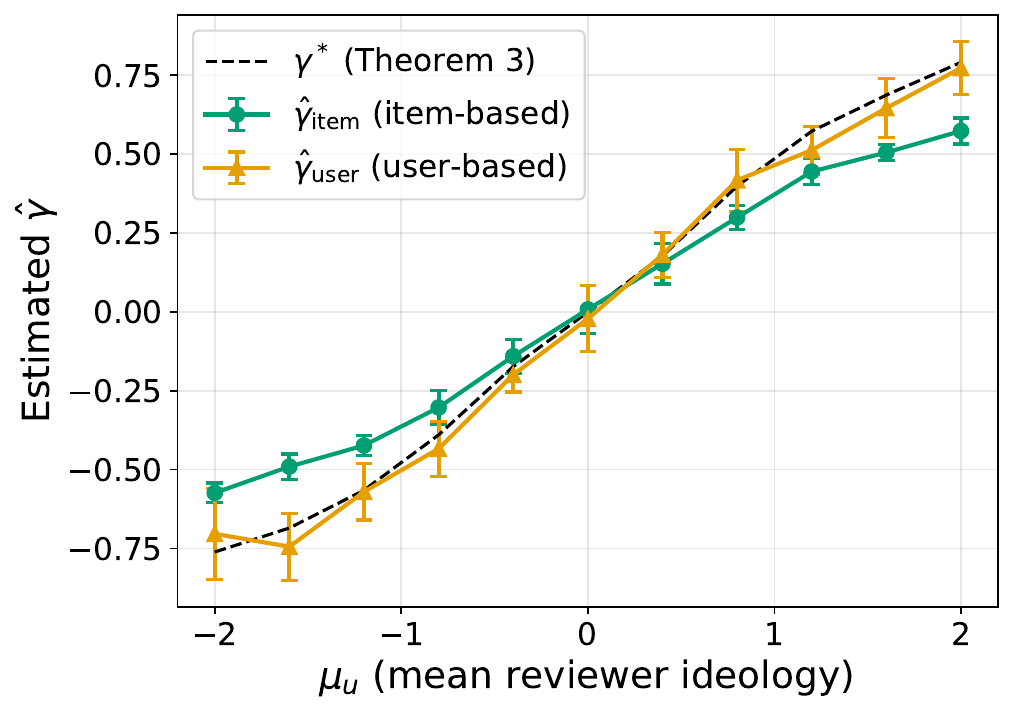}
    \caption{Estimated $\hat{\gamma}$ vs.\ $\mu_u$ under binary ratings (logistic link, Bernoulli observations, L2, $\kappa=1$). Dashed: $\gamma^*$ of \Cref{thm:anchor_l2}, a reference only, since the logistic link is outside the theorem. The user-based estimate tracks it; the item-based estimate compresses at large $|\mu_u|$, consistent with saturation of the logistic link.}
    \label{fig:gamma_binary}
\end{figure}

\begin{figure}[h]
    \centering
    \includegraphics[width=0.6\linewidth]{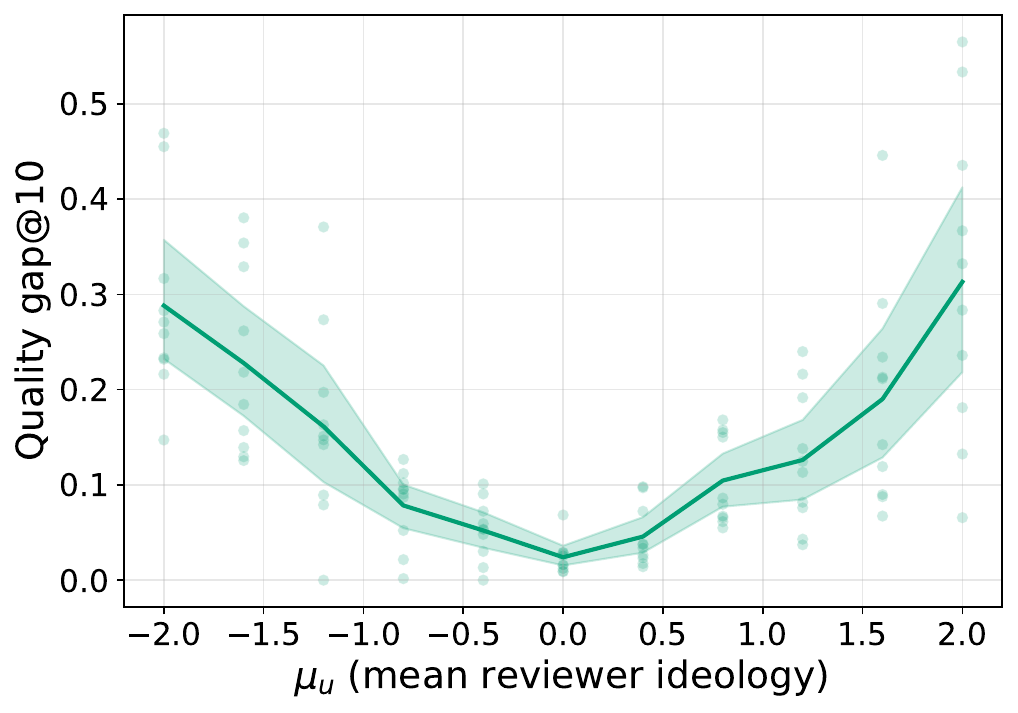}
    \caption{Quality gap@10 vs.\ $\mu_u$ under binary ratings. The monotonic increase with $|\mu_u|$ is preserved relative to the continuous case (\Cref{fig:quality_gap}).}
    \label{fig:oq_binary}
\end{figure}

Community Notes uses categorical (helpful, somewhat helpful, not helpful) ratings rather than continuous scores. To check that our results extend to this setting, we modify the DGP as follows: (i) compute the latent score $s_{ij} = q_j + v_j u_i + \varepsilon_{ij}$, (ii) pass it through the logistic link $\sigma(x) = 1/(1+e^{-x})$, and (iii) sample $r_{ij} \sim \mathrm{Bernoulli}(\sigma(s_{ij}))$. BMF is then fitted with binary cross-entropy loss; all other components of the procedure are unchanged. The $\gamma$--$\mu_u$ relationship is preserved, with a compression effect at large $|\mu_u|$ that is consistent with saturation of the logistic link. This indicates that the implicit anchoring to the population mean persists in the categorical setting actually used by Community Notes. Figures~\ref{fig:gamma_binary} and~\ref{fig:oq_binary} replicate the two main figures in this setting.

\subsection{Misspecified Rating Models}
\label{app:experiments:misspecification}

Our results assume ratings follow the bilinear form of \Cref{def:ground_truth}. To probe what survives when the generating model departs from it, we generate ratings as $r_{ij} = q_j + g(u_i, v_j) + \varepsilon_{ij}$ for three alternative interaction terms and fit the same $L_2$-regularized bilinear BMF in each case: a Euclidean utility $g = -(u_i - v_j)^2$, a higher-order interaction $g = u_i v_j + \varphi\,u_i^2 v_j^2$ with $\varphi = 0.3$, and an unmodeled confounder $g = u_i v_j + w_i z_j$ with $w_i, z_j \sim \mathcal{N}(0,1)$ drawn independently of the other parameters. The bilinear case $g = u_i v_j$ serves as the well-specified reference. Only the interaction term varies: every run uses the default DGP of \Cref{app:experiments:dgp} and the optimization settings of \Cref{app:experiments:implementation}. The three variants are specific departures from the bilinear form. Other misspecifications are untested.

\begin{figure}[h]
    \centering
    \includegraphics[width=\linewidth]{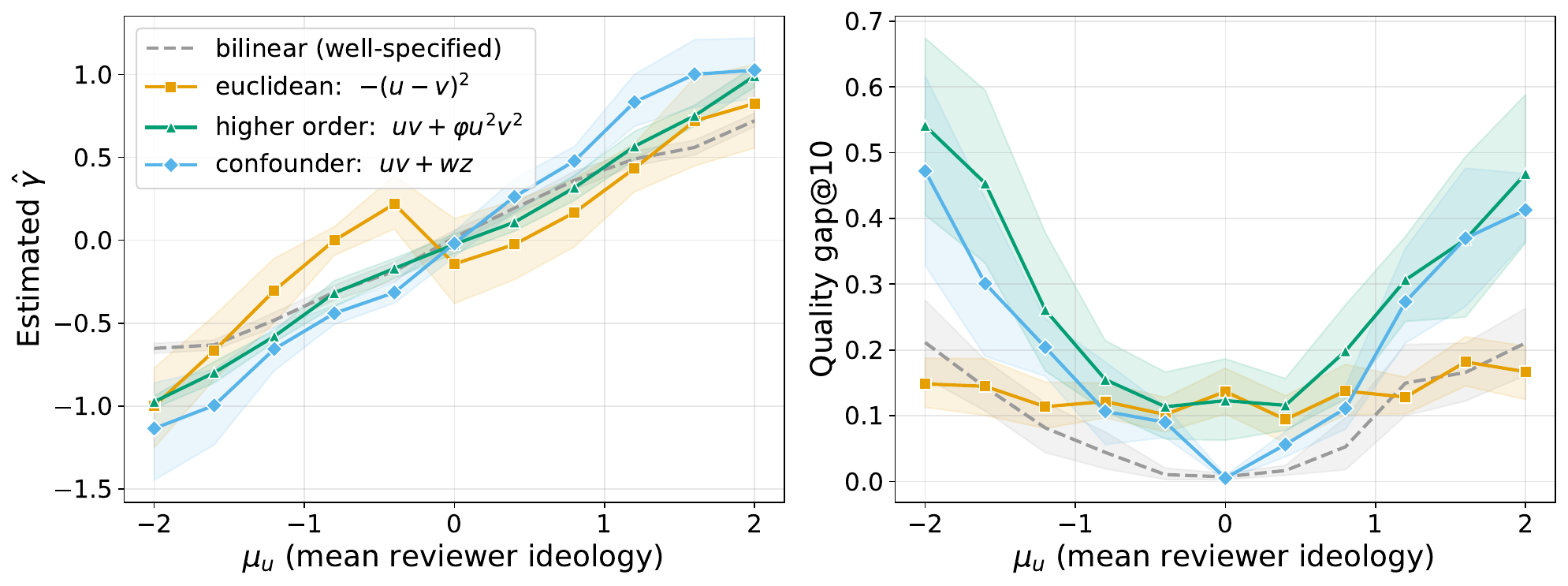}
    \caption{Estimated anchor $\hat{\gamma}$ (left) and Quality gap@10 (right) against mean reviewer ideology $\mu_u$, for the well-specified bilinear model and three misspecified variants. The anchor follows $\mu_u$ in all four; the quality gap vanishes at $\mu_u = 0$ for the bilinear model and the confounder, and stays positive under the Euclidean utility and the higher-order interaction. Bands are $\pm 1$ SE over 10 replications.}
    \label{fig:misspecification}
\end{figure}

\Cref{fig:misspecification} sweeps $\mu_u$ over $[-2, 2]$ for each variant. \Cref{tab:misspecification} reports the slope of each $\hat{\gamma}$ estimator against $\mu_u$ over that sweep, together with the quality gap at $\mu_u = 0$ and at $|\mu_u| = 2$.

The anchor tracks the mean reviewer ideology in every variant, with $\hat{\gamma}$ increasing in $\mu_u$ across the sweep. The slopes in \Cref{tab:misspecification} are fit across the whole range, where $\hat{\gamma}$ tapers at the extremes as it does under correct specification (\Cref{fig:gamma_sweep}), so they sit below the slope $1/2$ at $\mu_u = 0$; the exact curve of \eqref{eq:psi} has least-squares slope $0.42$ over this grid, close to the bilinear row ($0.37$, $0.38$), and the slopes are read as a comparison across variants.

Misspecification makes recovery worse. The confounder preserves recovery at $\mu_u = 0$, while the Euclidean utility and the higher-order interaction lose it, and the higher-order variant distorts most at the extremes. The two $\hat{\gamma}$ estimators, which agree under correct specification (\Cref{app:experiments:gamma_estimation}), also split under the Euclidean utility. The higher-order variant still takes its smallest gap at $\mu_u = 0$: a neutral pool remains the best case for quality recovery, though it no longer recovers the ranking exactly. Recovery at a neutral pool is proved for the bilinear model (\Cref{thm:anchor_l2}, \Cref{cor:quality_recovery}), while an anchor forms and follows the mean under every departure we ran.

\begin{table}[h]
    \centering
    \caption{Anchor slope and quality gap under misspecification. $s_{\mathrm{item}}$ and $s_{\mathrm{user}}$ are the slopes of the two $\hat{\gamma}$ estimators of \Cref{app:experiments:gamma_estimation} regressed on $\mu_u$ across the sweep of \Cref{fig:misspecification}. The two estimators agree under every variant except the Euclidean utility.}
    \label{tab:misspecification}
    % Slopes are least-squares fits over mu_u in [-2, 2].
% The exact anchor of Theorem 3 has slope 0.42 over this grid
% (1/2 is its slope at mu_u = 0), so the bilinear row is close to theory.
\begin{tabular}{llcccc}
\toprule
Variant & $g(u,v)$ & $s_{\mathrm{item}}$ & $s_{\mathrm{user}}$ & Quality gap$@10$ & Quality gap$@10$ \\
 & & & & ($\mu_u=0$) & ($|\mu_u|=2$) \\
\midrule
bilinear (well-specified) & $uv$ & 0.370 & 0.382 & 0.007 & 0.211 \\
euclidean & $-(u-v)^2$ & 0.386 & 0.193 & 0.137 & 0.158 \\
higher order & $uv + \varphi u^2 v^2$ & 0.478 & 0.500 & 0.123 & 0.504 \\
confounder & $uv + wz$ & 0.584 & 0.611 & 0.005 & 0.442 \\
\bottomrule
\end{tabular}

\end{table}

\subsection{Item Ideology Sweep}
\label{app:experiments:item}

\begin{figure}[h]
    \centering
    \includegraphics[width=0.6\linewidth]{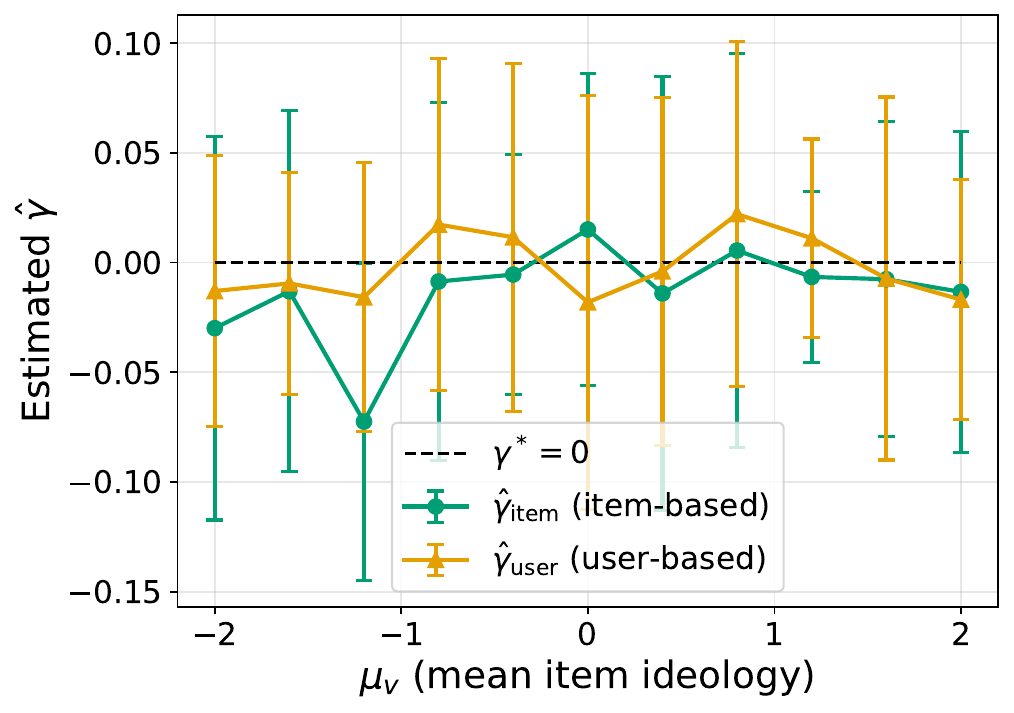}
    \caption{Estimated $\hat{\gamma}$ vs.\ mean item ideology $\mu_v$ (L2, $\kappa=1$, $\mu_u=0$). The theoretical prediction $\gamma^* = 0$ (dashed) is confirmed: item ideology mean does not shift the anchor.}
    \label{fig:gamma_item}
\end{figure}

\begin{figure}[h]
    \centering
    \includegraphics[width=0.6\linewidth]{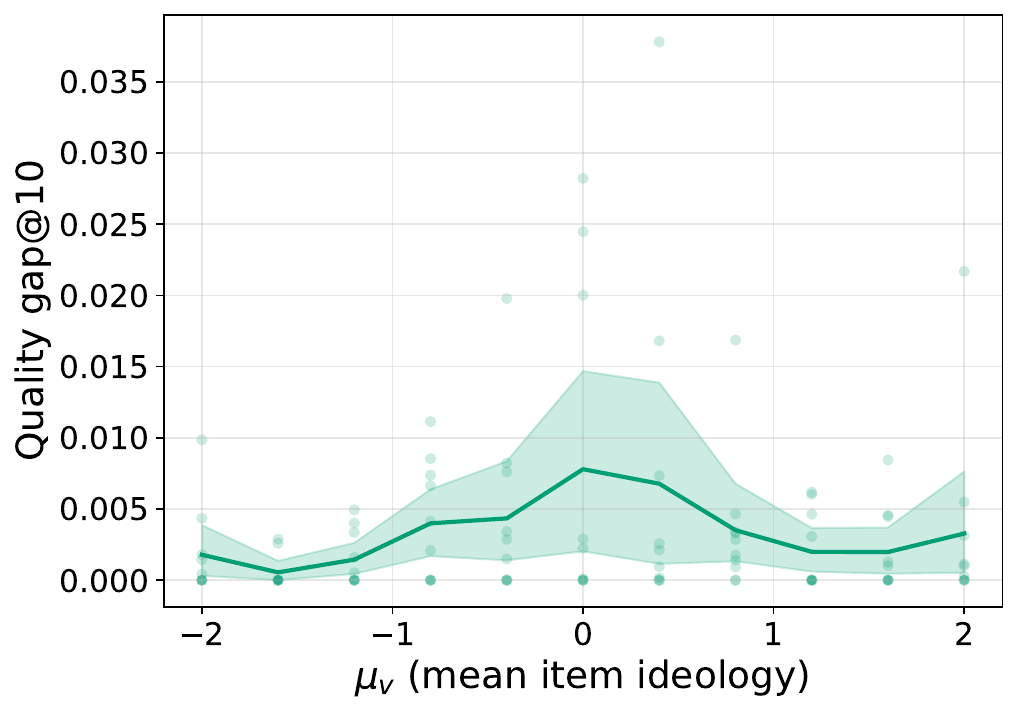}
    \caption{Quality gap@10 vs.\ mean item ideology $\mu_v$. Unlike the agent-mean sweep (\Cref{fig:quality_gap}), there is no systematic increase, confirming that the item ideology distribution does not induce quality distortion.}
    \label{fig:oq_item}
\end{figure}

Theory predicts that shifting the mean item ideology $\mu_v$ does not induce a gauge shift ($\gamma^* = 0$), since the anchor depends on the user distribution rather than the item distribution. (\Cref{thm:anchor_l2} assumes $\mu_v = 0$; the prediction still holds without it, since at $\mu_u = 0$ the intercept balance forces $\beta = 0$ even when $\eta^* \neq 0$, and the agent-centric balance then gives $\gamma = 0$.) We verify this by sweeping $\mu_v \in [-2, 2]$ while holding the user distribution fixed at $\mu_u = 0$ (all other parameters at their defaults). Both $\hat{\gamma}$ (\Cref{fig:gamma_item}) and the quality gap (\Cref{fig:oq_item}) remain flat across the full range of $\mu_v$, confirming that item ideology distribution is not a source of distortion.

\clearpage

\section{Use of Large Language Models}
\label{app:llm_usage}

Large language models (LLMs), primarily Claude (Anthropic), were used in the following supporting roles:

\begin{itemize}
    \item \textbf{Writing and editing:} improving clarity, grammar, 
    and flow; in some cases generated text was edited and incorporated.
    \item \textbf{Coding support:} refactoring experiment code, adding 
    evaluation metrics, and generating plotting utilities.
    \item \textbf{Proof assistance:} checking intermediate steps and 
    suggesting proof strategies. All theoretical results were developed 
    and verified by the authors.
    \item \textbf{Ideation:} brainstorming and exploratory discussion. 
    All conclusions are the authors' own.
\end{itemize}

LLMs did not contribute novel methods, theorems, or results to this work.

\end{document}